\documentclass[lettersize,journal]{IEEEtran}
\usepackage{amsmath,amsfonts}
\usepackage{algorithmic}
\usepackage{algorithm}
\usepackage{array}
\usepackage[caption=false,font=normalsize,labelfont=sf,textfont=sf]{subfig}
\usepackage{textcomp}
\usepackage{stfloats}
\usepackage{url}
\usepackage{verbatim}
\usepackage{graphicx}
\usepackage{cite}
\usepackage{gensymb}
\begin{document}

\title{Pix2HDR - A pixel-wise acquisition and deep learning-based synthesis approach for high-speed HDR videos}

\author{
    \IEEEauthorblockN{Caixin Wang\IEEEauthorrefmark{1}, Jie Zhang\IEEEauthorrefmark{1}, Matthew A. Wilson, Ralph Etienne-Cummings}

\thanks{C. Wang and R. Etienne-Cummings are with the Department of Electrical and Computer Engineering, The Johns Hopkins University, Baltimore, MD, USA}

\thanks{J. Zhang and M.A. Wilson are with the Picower Institute for Learning and Memory and the Department of Brain and Cognitive Sciences, Massachusetts Institute of Technology, Cambridge, MA, USA}

\thanks{\IEEEauthorrefmark{1} C. Wang and J. Zhang contributed equally}

}

\markboth{}%
{Shell \MakeLowercase{\textit{et al.}}: A Sample Article Using IEEEtran.cls for IEEE Journals}

\IEEEpubid{}

\maketitle

\begin{abstract}
Accurately capturing dynamic scenes with wide-ranging motion and light intensity is crucial for many vision applications. However, acquiring high-speed high dynamic range (HDR) video is challenging because the camera's frame rate restricts its dynamic range. Existing methods sacrifice speed to acquire multi-exposure frames. Yet, misaligned motion in these frames can still pose complications for HDR fusion algorithms, resulting in artifacts. Instead of frame-based exposures, we sample the videos using individual pixels at varying exposures and phase offsets. Implemented on a monochrome pixel-wise programmable image sensor, our sampling pattern captures fast motion at a high dynamic range. We then transform pixel-wise outputs into an HDR video using end-to-end learned weights from deep neural networks, achieving high spatiotemporal resolution with minimized motion blurring. We demonstrate aliasing-free HDR video acquisition at 1000 FPS, resolving fast motion under low-light conditions and against bright backgrounds — both challenging conditions for conventional cameras. By combining the versatility of pixel-wise sampling patterns with the strength of deep neural networks at decoding complex scenes, our method greatly enhances the vision system's adaptability and performance in dynamic conditions.
\end{abstract}


\section{Introduction}
\IEEEPARstart{D}{igital} cameras have become integral to many vision applications, including videography, smart devices, autonomous vehicles, and scientific research. The camera’s performance directly impacts these systems, and failure to capture crucial video features can have critical consequences. However, conventional cameras have limited dynamic range and temporal resolution at a fixed frame rate. They struggle to capture scenes with a large range of motion and intensities: fast events are susceptible to aliasing and blurring, high-intensity events cause overexposure, and low-light regions suffer from a low signal-to-noise ratio (SNR).

Extending the dynamic range for high-speed videos is fundamentally challenging, given the conventional camera’s inability to accommodate multiple exposures at each video frame. While multi-camera systems are commonly employed to boost dynamic range, they do so at the expense of large size, higher power consumption and complexity \cite{wilburn2005high, popovic2016multi, tocci2011versatile}. Alternatively, computational techniques can merge time-interleaved frames acquired from a single camera at both short and long exposures to extend the dynamic range \cite{kalantari2019deep, chen2021hdr}. But its performance heavily relies on motion alignment accuracy in under- and over-exposed regions and may introduce ghosting artifacts. Interpolation applied to frame pairs of short and long exposures can facilitate the alignment of inter-frame motion in scenarios involving slow motions \cite{khan2022deephs}. However, it is less effective in aligning and reconstructing high-speed motions, where significant changes occur between consecutive frames.

\begin{figure}[!t]
\centering
\includegraphics[width=3.49in]{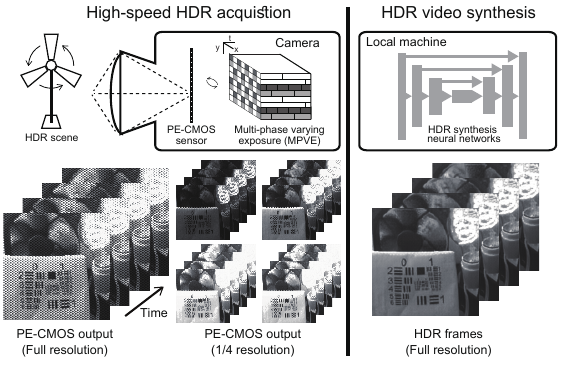}
\caption{The overview of the Pix2HDR acquisition / synthesis. Pix2HDR acquires the HDR scene using the multi-phase varying exposure (MPVE) pixel-wise sampling pattern, implemented on a CMOS image sensor (PE-CMOS). MPVE configures pixels into different exposures, speeds, and phase offsets to enhance temporal resolution and dynamic range. Pix2HDR synthesizes a high spatiotemporal HDR video from pixel-wise outputs using weights obtained through a deep neural network through end-to-end training, achieving high spatiotemporal resolution with minimized motion blurring.}
\label{fig_1}
\end{figure}

Recent innovations enabled computational imaging sensors and systems to control exposure at the pixel level \cite{martel2020neural, nayar2000high, nayar2006programmable, serrano2016convolutional, alghamdi2019reconfigurable, gulve202339, luo202230, hajisharif2015adaptive, zhang2020closed, zhang2016compact}. This allows the sensors to capture local high-light areas in an HDR scene with adjacent pixels configured in varying exposure durations. While the pixel-wise imaging configuration significantly simplifies cross-frame motion alignment, it comes at the expense of reduced spatial resolution \cite{nayar2000high}. Furthermore, the temporal resolution of these systems is limited by the longest-exposing pixels \cite{martel2020neural, luo202230, gulve202339}. While computational methods can recover video motion faster than the frame rate using Compressed Sensing principles, the reconstruction performance heavily depends on scene sparsity \cite{martel2020neural, luo202230, gulve202339, serrano2016convolutional, portz2013random}.

We present a pixel-wise imaging approach (Pix2HDR) for sampling HDR scenes at high-speed. Pix2HDR first acquires the scene using a multi-phase varying exposure pixel-wise exposure pattern (MPVE) (\textbf{Fig. \ref{fig_1}}). By using multiple pixels at different exposure durations and phase offsets, MPVE guarantees that each small pixel patch captures high-temporal resolution scene details with an extended dynamic range. Pix2HDR then transforms pixel-wise outputs into high-speed HDR videos using a deep neural network (LDR-HDR network) (\textbf{Fig. \ref{fig_1}}), leveraging learned weights to achieve high spatiotemporal resolution and dynamic range while eliminating aliasing and motion blur.

The rest of the paper is organized as follows: Section II provides a short summary of previous works. Section III offers a brief overview of our system, and in Section IV, we outline an optimized pixel-wise strategy for capturing HDR videos. Section V then delves into the specifics of Pix2HDR’s architecture for synthesizing high-speed HDR videos from pixel-wise outputs. For results in Section VI, we first evaluate our method’s performance against other works using publicly available datasets. We then capture HDR scenes using our method to assess its spatial resolution, temporal resolution, dynamic range, motion deblurring performance, and the capability to detect object motion under low-light conditions and against bright backgrounds. We conclude in Section VII by discussing its applications, greater impact, encoding scheme, system limitations, and future work.

\section{Previous work on HDR imaging}
\subsubsection{Acquiring and merging multiple frames of varying exposure} Learning-based or classic tone-mapping methods can merge time-synchronized low dynamic range (LDR) frames captured at different exposures into a single HDR frame \cite{mitsunaga1999radiometric, debevec2023recovering, reinhard2010high, gallo2009artifact, sen2012robust, oh2014robust, kalantari2017deep, cai2018learning, yan2020deep, yao2011robust, baker2011database, zimmer2011freehand}. However, the performance of these methods depends on additional alignment techniques to rectify differences between frames due to motion, overexposure, and occlusions. Failure to synchronize the LDR frames results in ghosting artifacts in the merged HDR frame. 

Obtaining time-aligned frames can also be challenging and often requires using multiple cameras. These setups either split incoming light using beam splitters among different cameras \cite{mcguire2007optical, tocci2011versatile, froehlich2014creating}, or employing an array of cameras, each with its independent optical path \cite{wilburn2005high, ramachandra2008hdr, popovic2016multi}. Multi-camera systems add significant complexity, bulkiness, and increased power consumption.  Moreover, dividing the incoming light among multiple detectors degrades the overall SNR. Additionally, camera arrays demand parallax correction to align cameras’ FOV, which leaves room for artifacts and errors.

\subsubsection{HDR imaging through pixel-wise exposure modulation} Pixel-wise exposure modulation allows precise control of each pixel’s exposure duration. Given natural scenes’ inherent high spatial correlation, local pixels, set with distinct exposures, can capture the scene at varying dynamic ranges. This method simplifies the alignment between multiple exposure pixels and allows HDR reconstruction using simple spatial filtering (or learning-based methods) but at the cost of lower spatial resolution \cite{nayar2000high, nayar2003adaptive, nayar2006programmable, martel2020neural, serrano2016convolutional, alghamdi2019reconfigurable, hajisharif2015adaptive}.

Previously implemented sensors with pixel-wise exposure control have several key disadvantages: exposure control circuits at the pixel level occupy valuable detector areas and reduce the photodiode fill factor \cite{martel2020neural, carey2013100}. This degrades the pixel sensitivity compared to a conventional CMOS image sensor. Furthermore, previous pixel-wise exposure sensors have slow frame rates ($\sim$30 FPS), constrained by the longest pixel exposure duration \cite{nayar2000high, nayar2003adaptive, nayar2006programmable, martel2020neural, jiang2021hdr, cogalan2022learning}. This leads to pixels with shorter exposures operating at the same sampling rate as those with longer exposures and limits the sensor’s ability to perform high-speed imaging.

\section{Overview}
The Pix2HDR consists of an acquisition system and a synthesis method (\textbf{Fig. \ref{fig_1}}). The acquisition system is a camera built with the pixel-wise programmable exposure (PE-CMOS) image sensor\cite{zhang2023pixel}. Individual pixels in PE-CMOS can be separately exposed. This feature allows for different exposure durations and phases at each pixel, making it possible to coordinate their configurations to optimize the temporal resolution and dynamic range during high-speed imaging. PE-CMOS features an efficient pixel design and achieves a 75\% photodiode fill factor with a 10 \textmu m pixel pitch. The PE-CMOS has comparable low-light performance to state-of-the-art low-noise CMOS sensors without pixel-wise exposure control\cite{zhang2023pixel}.

Using the PE-CMOS sensor, we implemented the MPVE sampling pattern optimized for high-speed HDR imaging. In MPVE, neighboring pixels have variable exposure, sampling rate, and phase offsets (\textbf{Fig. \ref{fig_1}}). As we will demonstrate in Section IV, varying pixel-wise exposure expands the dynamic range and prevents motion blur. Relative phase offsets in adjacent pixels enhance the acquisition’s temporal resolution without increasing pixel sampling speed. We will show that with a 250 Hz pixel sampling rate, the Pix2HDR method can sample the scene with a 1000 Hz temporal resolution. 

The Pix2HDR video synthesis algorithm employs deep neural networks to determine the optimal weights for converting the sensor’s pixel outputs into high-speed HDR video. The deep learning-based method efficiently addresses the conflicting demands of high spatiotemporal resolution and dynamic range while preventing aliasing and motion blur. As described in Section V, the Pix2HDR synthesis networks, implemented on an Nvidia RTX3080, have achieved an average frame inference time of 2.5 ms and support a 400 FPS real-time HDR video synthesis.

\begin{figure}[!t]
\centering
\includegraphics[width=3.49in]{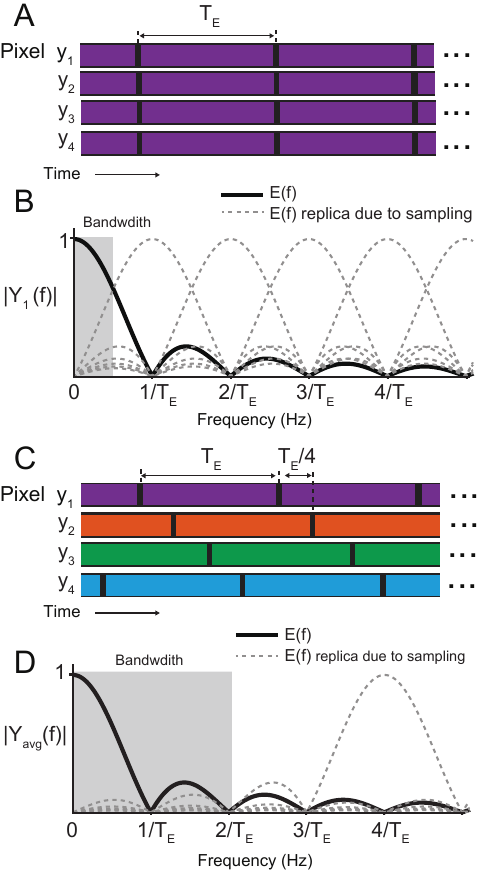}%
\caption{Multi-phase sampling enhances the temporal resolution without increasing sampling speed. \textbf{A.} Conventional camera samples all the pixels ($y_1,...,y_4$) concurrently (global shutter) or in fast line sequences (rolling shutter) with pixel exposure of $T_E$ and sampling rate at $1/T_E$. \textbf{B.} Frequency spectrum of the averaged pixel value, $y_{\rm avg}$. Its Nyquist bandwidth is limited to $1/(2T_E)$ and suffers from a significant amount of temporal aliasing. \textbf{C.} In multi-phase sampling, pixels of exposure $T_E$ are phase-offsetted in multiples of $T_E/4$. \textbf{D.} Without increasing the sampling rate, multi-phase exposures extend the $y_{\rm avg}$ bandwidth by four times to $2/T_E$ and mitigate temporal aliasing by pushing the replica spectra to higher frequencies. }
\label{fig_2}
\end{figure}

\section{Multi-phase varying exposure (MPVE) pixel-wise sampling configuration}

We proposed the MPVE pixel-wise sampling configuration to maximize temporal resolution, SNR (for fast transient events and static scenes) and dynamic range at sampling HDR scenes. Here we discuss each consideration and derive the MPVE sampling pattern for HDR video acquisition. 

\subsection{Eliminating temporal aliasing with pixel-wise phase offset}
The camera’s temporal resolution is an essential consideration for high-speed imaging. The selection of pixel exposure, $T_E$, establishes the pixel's sampling rate to $1/T_E$, along with its Nyquist bandwidth at $1/(2T_E)$. For a conventional camera, where pixels are either all simultaneously integrated (global shutter) or sequentially in rows (rolling shutter), the choice of $T_E$ also limits the temporal resolution of the camera (\textbf{Fig. \ref{fig_2}A}). We can increase the temporal resolution by shortening $T_E$. But it comes at the cost of lower SNR and higher sampling speed and higher power consumption. 

While we cannot increase the temporal resolution of a single pixel without reducing $T_E$, we could exploit the phase relationship between multiple pixels to enhance the sensor’s temporal resolution. This approach is viable due to the inherent spatial correlations of natural scenes, where closely situated pixels capture closely related activities. To illustrate (\textbf{Fig. \ref{fig_2}}), we assume a light impulse, $\delta(t)$, is simultaneously sampled by 4 adjacent pixels, $y_1(n)$, ..., $y_4(n)$, each with an exposure function $e(t)$. The single pixel output $y_1(t)$ is:
\begin{equation}
\label{eq1}
y_1(t)=e(t)*\delta(t),\quad
e(t)=
\begin{cases}
    1,&{0 \leq t \leq T_E},\\
    {0,} &{\text{otherwise,}}
\end{cases} 
\end{equation}

\noindent $y_1(t)$ is then sampled by an ADC at a period of $T_E$, the discretized version, $y_1[n]$ is:

\begin{equation}
\label{eq2}
y[n]=y(nT_E),
\end{equation}

\noindent with frequency spectrum:

\begin{equation}
\label{eq3}
Y_1(f)=\frac{1}{T_E} \sum_{n=-\infty}^{\infty} E(f-\frac{n}{T_E}),
\end{equation}

\noindent where $E(f-n/T_E)$ are the replicas of the exposure function’s spectrum resulting from sampling. We can plot $Y_{1}(f)$ to see that the majority of the frequency spectrum is aliased, resulting in signal distortion (\textbf{Fig. \ref{fig_2}B}).

By relying on the phase difference between neighboring pixels, we can eliminate the aliasing effect without increasing each pixel’s sampling speed. To achieve this, relative to the phase of pixel 1, we can introduce phase shifts of $T_E/4$, $T_E/2$, $(3T_E)/4$ to pixel 2, 3, and 4 respectively (\textbf{Fig. \ref{fig_2}C}). These pixels’ spectrum, $Y_{k}(f)$, becomes: 
\begin{equation}
\label{eq4}
Y_{k}(f)=\frac{1}{T_E} \sum_{n=-\infty}^{\infty} E(f-\frac{n}{T_E})e^{-j2\pi\frac{n(k-1)}{4}},
\end{equation}

\noindent where $k\in\{1,2,3,4\}$. If we average these four pixels, the resulting spectrum becomes: 
\begin{equation}
\label{eq5}
\begin{split}
Y_{\rm avg}(f)&=\frac{1}{4} \sum_{k=1}^{4}Y_{k}(f)=\\
&\frac{1}{4T_E}\sum_{n=-\infty}^{\infty}E(f-\frac{n}{T_E})\sum_{k=1}^{4}e^{-j2\pi\frac{n(k-1)}{4}},
\end{split}
\end{equation}

\noindent since $Y_{\rm avg}(f)=0$ when $n$ is not a multiple of 4, the above equation can be rewritten as:
\begin{equation}
\label{eq6}
Y_{\rm{avg}}(f)=\frac{1}{4T_E} \sum_{n=-\infty}^{\infty}E(f-\frac{4n}{T_E}),
\end{equation}

\noindent which spaces the frequency replicas of $E(f)$ away from each other and avoids aliasing (\textbf{Fig. \ref{fig_2}D}). The average spectrum of phase shifted pixels increases the Nyquist bandwidth of individual pixels by four times, without increasing in per-pixel sampling rate. From a time-domain viewpoint, the combined phase-shifted pixel outputs are equivalent to the sampling of the impulse signal, $\delta(t)$, at a higher rate of frequency, $4/T_E$. We will demonstrate this using an experiment in Section VI (\textbf{Fig. \ref{fig_10}})

\subsection{Maximizing SNR for high-speed signals} 
Maximizing the SNR for detecting ultra-fast events is another important consideration for the pixel sampling pattern. Fast events can result from various phenomena, such as the fast fluorescent intensity change in biological tissues (neurons, cardiac cells, etc.) and the rapid movement of small objects, whose signals appear as pulsed events at individual pixels. The optimal duration of exposure that maximizes the SNR for detecting these events depends on the temporal characteristics of the signal. The SNR may be compromised if pixels are sampled with an inappropriate exposure time: A short exposure time is inadequate for pixel to integrate the signal, while an excessively long exposure time increases shot noise without increasing signal power, resulting in SNR degradation.

\begin{figure}[!t]
\centering
\includegraphics[width=3.49in]{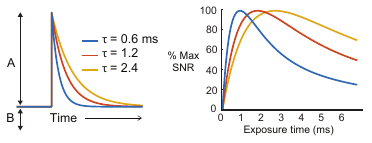}
\caption{Exposure duration determines pixel SNR at sampling high-speed events. To illustrate: \textbf{Left.} Transient events of amplitude A with baseline intensity B, and different falling time constants ($\tau$) of 0.6, 1.2 and 2.4 ms. \textbf{Right.} Pixel SNR with respect to the exposure time, $T_E$, for the signals on the left. Each curve is normalized with the maximum SNR set to 100\%. The SNR increases with longer exposure time but drops as extra integration time adds more shot noise than signal power.}
\label{fig_3_snr}
\end{figure}

We can demonstrate this by deriving an expression for a pixel's SNR. Let $v(t)$ represent a fast transient signal at a particular pixel. This could result from the movement of an ultra-fast object or the fast change of light intensity of a stationary object. For simplicity, $v(t)$ is denoted as a signal with an instantaneous rise time and a fall time modeled by an exponential function with time constant, $\tau$, amplitude $A$, and baseline intensity $B$ (\textbf{Fig. \ref{fig_3_snr}}):
\begin{equation}
\label{eq7}
v(t)=
\begin{cases}
    B,&{\text{when}}\ t<0,\\
    B+Ae^{-t/\tau},&{\text{when}}\ t\geq 0.
\end{cases}
\end{equation}

Using Eq.\ref{eq1}, the pixel output $y(t)$ is obtained as the convolution result between $v(t)$ and the exposure $e(t)$. The discrete samples, $y[n]$, of the resulting output are written as:
\begin{equation}
\label{eq8}
y[n]=y(nT_E)+\sigma[n],
\end{equation}

\noindent where $\sigma [n]$ is the noise containing both shot noise and circuit read noise. The shot noise is described by a Poisson distribution with a rate factor $\lambda$ equal to the signal at the pixel. As such, the noise power is expressed as:
\begin{equation}
\label{eq9}
N_{\rm shot}=y[n]+B\cdot T_E,
\end{equation}

If we assume the pixel is operating in a shot noise-limited region, where the read noise is negligible (i.e. $B\gg N_{\rm read}$). The SNR can be expressed as: 
\begin{equation}
\label{eq10}
\text{SNR}=\frac{S}{N_{\rm shot}}=\frac{{\rm max}\{y\}^2}{{\rm max}\{y\}+B\cdot T_E}.
\end{equation}

We plotted the SNR with respect to $T_E$ for signals of different time constants (\textbf{Fig. \ref{fig_3_snr}}). The SNR curves exhibit a concave pattern in relation to $T_E$: SNR first grows as the increase in $T_E$ allows more signal energy to be integrated into the pixel. However, once a peak value is reached, further increase of $T_E$ no longer results in additional energy integration but raises the shot noise power due to the baseline intensity, $B$. 

\begin{figure}[!t]
\centering
\includegraphics[width=3.49in]{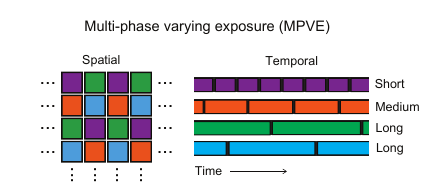}
\caption{MPVE sampling pattern uses varying exposures (short, medium, and long) at different phase offsets to maximize SNR for signals of different temporal characteristics.}
\label{fig_3_mpve}
\end{figure}

\begin{figure}[!t]
\centering
\includegraphics[width=3.3in]{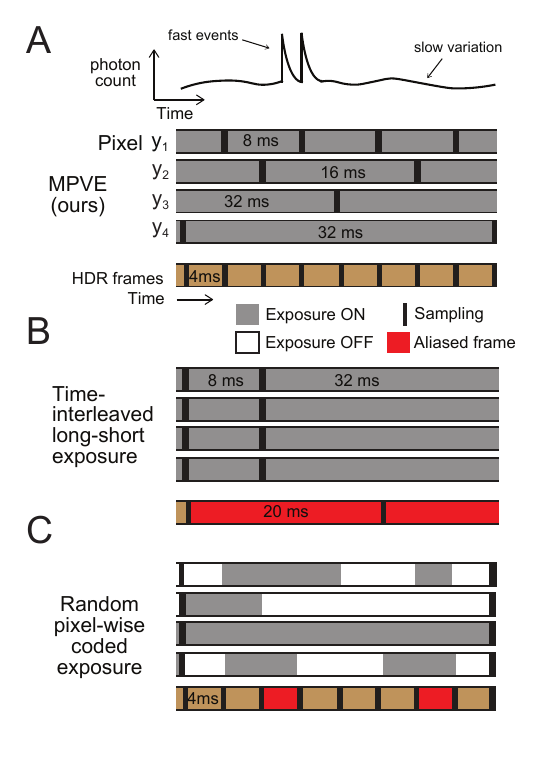}
\caption{Temporal comparison of MPVE vs other sampling patterns. \textbf{A.} Sampling a signal that consists of a slow variation with two closely spaced fast events, the MPVE guarantees aliasing-free sampling at high time resolution. Long pixel exposure also increases the SNR of the slow variations and expands the dynamic range. \textbf{B.} Sampling the same signal using time-interleaved long-short exposure will introduce aliasing, unable to resolve the timing of these two events. \textbf{C.} Multiple-on coded exposure pattern distributed randomly also creates aliasing: events appearing at either of the red colored frames generate the same reading at all pixel outputs, creating ambiguity in the events’ timing.}
\label{fig_4}
\end{figure}

\begin{figure}[!t]
\centering
\includegraphics[width=3.49in]{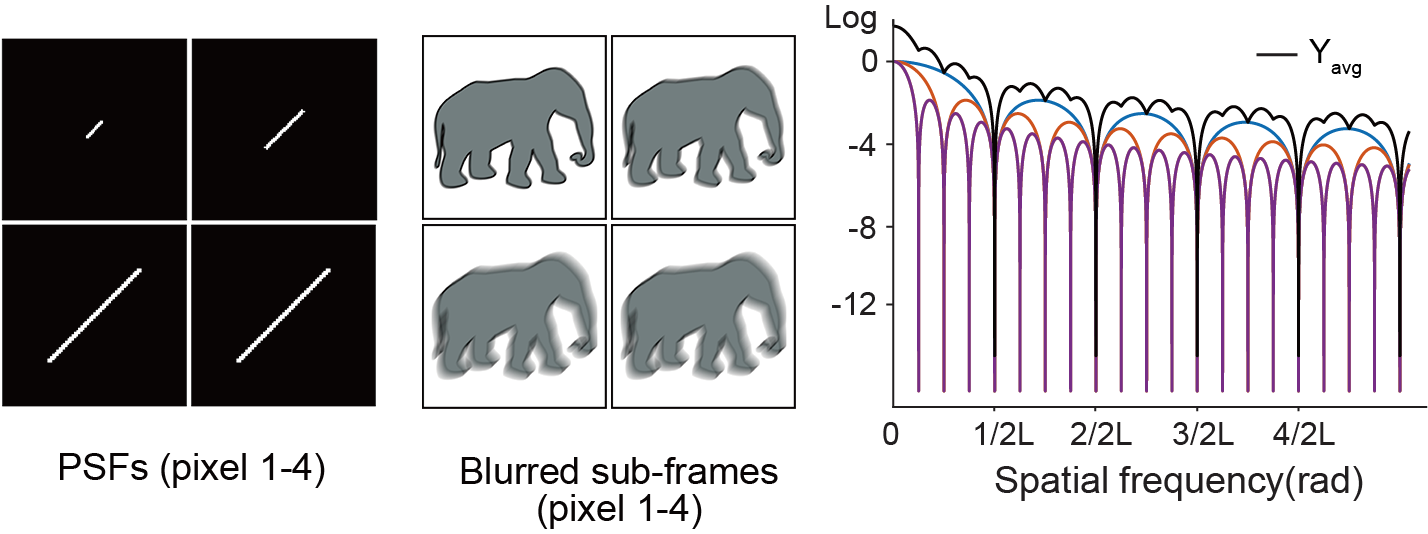}
\caption{PSF kernels of the MPVE exposure pattern. The same motion translates to PSF length $2L$ to $8L$ for at pixel $y_1$ to $y_4$, depending on their exposure. The corresponding spectra are sinc functions with zeros at multiples of $1/2L$, $1/4L$, $1/8L$, causing unrecoverable blurring effects. The spectrum of MPVE output, $Y_{\rm avg}$, eliminates zeros except at $1/2L$, achieving deblurring goal of which means the blurring effect of ideal recovered frame keeps at $2L$. }
\label{fig_5}
\end{figure}

The relationship of SNR with $T_E$ indicates that a continuous exposure function of appropriate duration is the key to maximizing SNR. To improve the SNR of signals encompassing a range of frequencies, we can configure adjacent pixels to have different exposure durations (\textbf{Fig. \ref{fig_3_mpve}}). This leads to the Multi-Phase Varying Exposure (MPVE) pixel-wise sampling pattern. Every $2\times2$ patches are arranged to have a short ($T_E/2$), medium ($T_E$) and long exposure ($2T_E$) arranged with relative phase offset of $T_E/4$, reaching the overall temporal resolution of $T_E/4$.

We can qualitatively illustrate the MPVE’s advantages over other patterns at sampling a signal with diverse temporal characteristics (\textbf{Fig. \ref{fig_4}A}). The signal, common to four pixels ($y_1, ...,y_4$), comprises a faint, slowly fluctuating baseline with two rapid burst events occurring 4 ms apart. MPVE’s phase-shifted pixel arrangement ensures that fast events’ timing can be unambiguously acquired (\textbf{Fig. \ref{fig_4}A}). In addition, short exposures are used to maximize the SNR of the burst events. At the same time, the medium and long exposure at $y_2$, $y_3$ and $y_4$ enhances the SNR of the weak baseline signal. 

In contrast to MPVE, the time-interleaved long-short exposure method introduces aliasing in frames when burst events coincide with a long exposure periods (\textbf{Fig. \ref{fig_4}B}). The long exposure also limits the maximum frame rate. Similarly, in pixel-wise coded exposure, the random arrangement of exposure times can create ambiguity when trying to resolve the timing of fast events: the pixel outputs will be identical if a spike occur at the two alised frames, causing ambiguity (\textbf{Fig. \ref{fig_4}C}). Furthermore, this approach results in lower pixel SNR because photons are not sampled during exposure off-times.

\begin{figure*}[!t]
\centering
\includegraphics[width=6.5in]{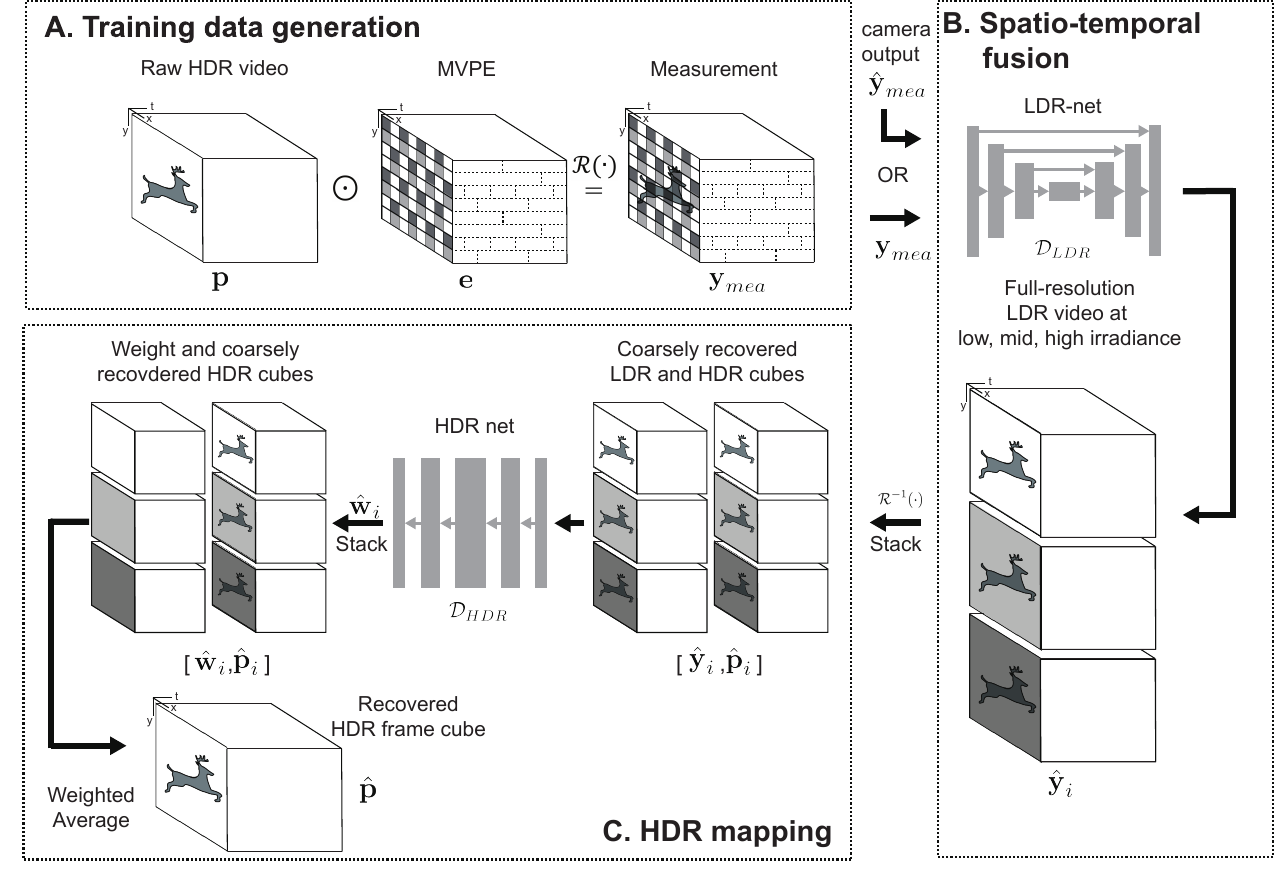}
\caption{LDR-HDR network for HDR video synthesis. \textbf{A.} Training data generation: The ground truth video, $\textbf{p}$, are modulated by MVPE cube, $\textbf{e}$, and then converted from irradiance value to camera outputs, $\textbf{y}_{\rm mea}$, through a camera response function (CRF), $\mathcal{R}(\cdot)$. \textbf{B.} Spatio-temporal fusion using LDR-net: $\textbf{y}_{\rm mea}$ (for training) or $\hat{\textbf{y}}_{\rm mea}$  (for testing) are inputs into the LDR-net $\mathcal{D}_{\rm LDR}(\cdot)$, which performs spatio-temporal up-sampling and renders three LDR video cubes $\hat{\textbf{y}}_i$ at low, medium, and high exposure levels. \textbf{C.} HDR mapping using HDR-net: $\hat{\textbf{y}}_i$ are first mapped to their corresponding irradiance, $\hat{\textbf{p}}_i$, using the inverse CRF, $\mathcal{R}^{-1}(\cdot)$. Then $\hat{\textbf{y}}_i$ and $\hat{\textbf{p}}_i$ are inputs into the HDR-net $\mathcal{D}_{\rm HDR}(\cdot)$ to obtain the weights $\hat{\textbf{w}}_i$ to fuse $\hat{\textbf{p}}_i$ into a single HDR video, $\hat{\textbf{p}}$.}
\label{fig_6}
\end{figure*}

\subsection{Minimizing blurring and extending the dynamic range}
The MPVE naturally extend the sampled dynamic range within a $2\times2$ pixel patch, as pixels acquire the scene irradiance at different exposures. In addition, The varying pixel exposures also suppress blurring that would have occurred if all pixels were exposed for the same duration. To illustrate, a motion across the FOV translates to point spread function (PSF) of length $2L$ to $8L$ at pixel $y_1$ to $y_4$, depending on their exposure length (\textbf{Fig. \ref{fig_5}}). These PSF’s spectra contain zeros that cause unrecoverable blurring effects \cite{agrawal2009invertible}. The PSF of the averaged pixel values in MPVE, $y_{\rm avg}$, can eliminate the zeros except at locations multiples of $1/2L$, reaching a deblurring resolution of $2L$. Incorporating non-integer multiples of pixel exposures can improve deblurring performance. However, its implementation might require adopting an alternative spatiotemporal pixel configuration, with distinct trade-offs in anti-aliasing and dynamic range.

\section{LDR-HDR networks for synthesizing HDR videos from pixel-wise outputs.}
With MPVE, each small pixel patch of PE-CMOS captures high-temporal resolution scene details with an extended dynamic range. The proposed video synthesis algorithm aims to convert pixel outputs into a high-speed HDR video, achieving high spatiotemporal resolution and dynamic range while eliminating aliasing and motion blur. Conventional linear models, which derive each pixel value through a linear combination of its neighboring pixels, often fail to meet all specifications simultaneously. Instead, we leverage a deep learning approach to find the optimal weights for transforming pixel-wise outputs into high spatiotemporal resolution HDR video. 

Our proposed LDR-HDR architecture consists of a pair of 3D CNN networks: LDR-net and HDR-net. LDR-net handles spatiotemporal upsampling/interpolation by transforming PE-CMOS outputs into three videos, $[\hat{\textbf{y}}_{\rm low}, \hat{\textbf{y}}_{\rm mid}, \hat{\textbf{y}}_{\rm high}]$ (\textbf{Fig. \ref{fig_6}}). These videos of lower dynamic range (LDR), with a spatial resolution of $Nc \times Nr$ and a temporal resolution of $Nt$, render the scenes at low, medium, and high irradiance levels. Meanwhile, HDR-net is responsible for HDR fusion, combining high-speed, high-resolution, low dynamic range video stacks into a single tone-mapped HDR video (\textbf{Fig. \ref{fig_6}}). Here we discuss the details of training data generation and the architecture of LDR-HDR network. 

\subsection{Training data generation: transforming ground truth HDR video to pixel outputs}

\begin{figure*}[!t]
\centering
\includegraphics[width=5.5in]{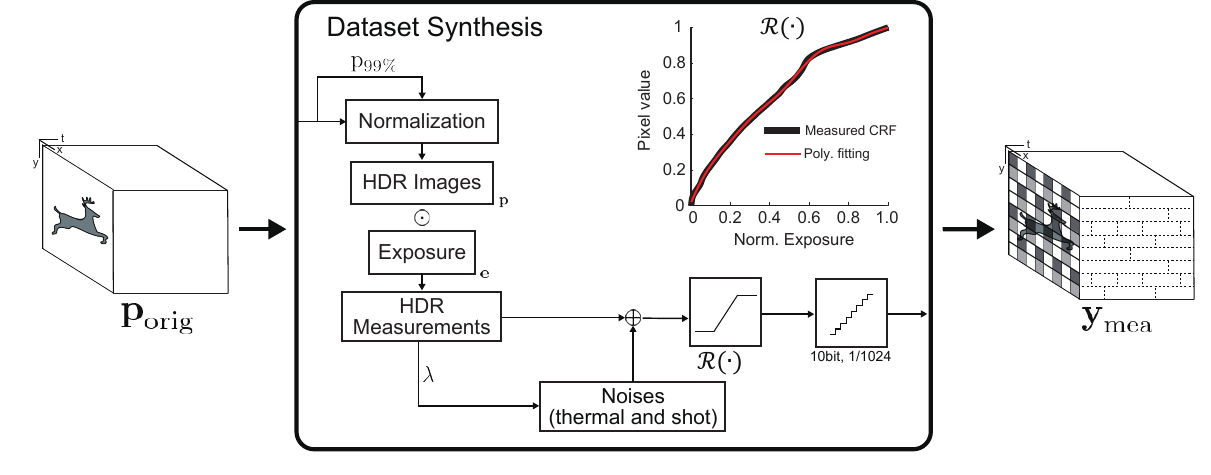}
\caption{Training data preparation. The training dataset, $\textbf{p}_{\rm orig}$, are firstly normalized with anchor irradiance value $\rm p_{99\%}$ to get $\textbf{p}$. We then obtain pixel output: $\textbf{y}_{\rm mea}=\mathcal{R}(\textbf{p}\odot \textbf{e}+\textbf{n}_{\rm thermal}+\textbf{n}_{\rm shot} )$. $\mathcal{R}(\cdot)$ is the measured camera CRF function. $\textbf{n}_{\rm shot}$ is the simulated shot noise with $\lambda$ =  $\textbf{p}\odot \textbf{e}$ .  $\textbf{n}_{\rm thermal}$ follows zero mean gaussian distribution with s.t.d set to the measured camera rms thermal noise. $\textbf{y}_{\rm mea}$ is quantized to 10bits to reflect PE-CMOS outputs resolution.}
\label{fig_7}
\end{figure*}

We generate training data using public HDR video datasets (hdm-hdr-2014) \cite{froehlich2014creating}. This process is summarized in \textbf{Fig. \ref{fig_6}} and expanded into details in \textbf{Fig. \ref{fig_7}}. Original HDR video cube, $\textbf{p}_{\rm orig}\in \mathbb{R}^{N_r \times N_c \times N_t}$, is first normalized by clipping the 1\% intensity outliers to ${\rm p}_{99\%}$, the 99\% percentile of the intensity range: 
\begin{equation}
\label{eq11}
\textbf{p}={\rm clip}(\frac{\textbf{p}_{\rm orig}}{{\rm p}_{99\%}}),
\end{equation}

\noindent where, $\textbf{p} \in \mathbb{R}^{N_r \times N_c \times N_t}$, is ground truth video and ${\rm clip}(\cdot)$ operations set any pixel value that is outside the range [0,1] to 1. 

$\textbf{p}$ is then transformed into pixel measurements, $\textbf{y}_{\rm mea} \in \mathbb{R}^{N_r \times N_c \times N_t}$ through modulation with the exposure cube, $\textbf{e}$:
\begin{equation}
\label{eq12}
\textbf{y}_{\rm mea}=\mathcal{R}(\textbf{p}\odot\textbf{e}+\textbf{n}).
\end{equation}

Here, the operator $\odot$ represents the PE-CMOS camera’s operation of pixel integration and sampling (\textbf{Fig. \ref{fig_6}, \ref{fig_7}}). We ensure $\textbf{y}_{\rm mea}$ has the same dimension as $\textbf{p}$ by up-sampling the resulting pixel values in time. A noise term, $\textbf{n}$, consisting of read noise and shot components, is then added to the pixel values. $\mathcal{R}(\cdot)$ is the measured camera response function (CRF) that corrects for the camera’s non-linear response between exposure and sampled pixel values (\textbf{Fig. \ref{fig_7}}). Finally, $\textbf{y}_{\rm mea}$ is quantized to a 10-bit resolution. The pairs of $\textbf{y}_{\rm mea}$ and $\textbf{p}$ form the training data of the LDR-HDR networks (LDR-net and HDR-net).

\subsection{LDR-net: Mapping pixel-wise outputs to high spatio-temporal resolution videos at different exposure levels}
\subsubsection{LDR-net architecture}
The goal of the LDR-net, denoted as $\mathcal{D}_{\rm LDR}$, is to transform pixel-wise outputs, $\hat{\textbf{y}}_{\rm mea}$, containing pixel outputs at different exposure, speed, and phase offsets, into three high-speed, high-spatial-resolution videos, $\hat{\textbf{y}}_{\rm low}, \hat{\textbf{y}}_{\rm mid}, \hat{\textbf{y}}_{\rm high}$, at low, medium and high exposure level. We adopted a 3D U-net architecture to accomplish this task: $\mathcal{D}_{\rm LDR}$ consists of an encoder module, a bottleneck module, and a decoder module. Three skip connections are attached between the encoder and decoder modules. Instead of using pooling layers, we adopt convolutional and transposed convolutional layers for down- and up-sample feature mapping. 

The encoder module has three blocks: the first block consists of a $7\times7\times7$ convolutional layer followed by a leaky rectified linear unit (LeakyReLU) and a batch normalization layer. In the other two blocks, the convolutional layer is replaced by two $3\times3\times3$ convolutional layers. While the stride of the first layer is 1, the stride length of the second layer rises to 2 to downsample the feature map size. 

The bottleneck module contains 12 residual blocks, where each block has two $3\times3\times3$ convolutional layers, between which there are a LeakyReLU, a batch normalization layer, and a skip connection. 

The decoder module contains three blocks, each of which contains a $4\times4\times4$ transposed convolutional layer with a stride equals to 2 and padding size equal to 1 (up-sampling layer), a $3\times3\times3$ convolutional layer, a LeakyReLU, and a batch normalization layer. The output video is non-linearized with a sigmoid function so that its range is between [0,1]. All operations are conducted in 4D tensors, and the input channel size is 1 (grayscale), and the output channel size is 3 (three LDR videos under low, mid, and high exposures).

\subsubsection{training}
$\mathcal{D}_{\rm LDR}$ learns the mapping for interpolating  $\textbf{y}_{\rm mea}$ to high spatiotemporal resolution LDR videos $[\hat{\textbf{y}}_{\rm low}, \hat{\textbf{y}}_{\rm mid}, \hat{\textbf{y}}_{\rm high}]$:
\begin{equation}
\label{eq13}
[\hat{\textbf{y}}_{\rm low}, \hat{\textbf{y}}_{\rm mid}, \hat{\textbf{y}}_{\rm high}]=\mathcal{D}_{\rm LDR}(\textbf{y}_{\rm mea}).
\end{equation}

To train the $\mathcal{D}_{\rm LDR}$’s parameters, we first generate ground truth LDR videos from raw HDR video $\textbf{p}$:
\begin{equation}
\label{eq14}
\textbf{y}_{\rm low}=\mathcal{R}(2\textbf{p}), \quad \textbf{y}_{\rm mid}=\mathcal{R}(4\textbf{p}), \quad \textbf{y}_{\rm high}=\mathcal{R}(8\textbf{p}).
\end{equation}

The training process minimize the loss function $\mathcal{L}(\cdot)$ with the ADAM stochastic gradient-based optimization algorithm \cite{kingma2014adam}:
\begin{equation}
\label{eq15}
\mathop{\arg \min}_{\mathcal{D}_{\rm LDR}}\sum_{i}\mathcal{L}(\hat{\textbf{y}}_i, \textbf{y}_i),
\end{equation}

\noindent where $i\in\{{\rm low, mid, high}\}$ and $\mathcal{L}(\cdot)$ is a combination of three loss functions, defined as:
\begin{equation}
\label{eq16}
\mathcal{L}=\mathcal{L}_{\rm MSE} + \mathcal{L}_{\rm perc} + \mathcal{L}_{\rm SSIM}.
\end{equation}

\subsubsection{Loss function}
$\mathcal{L}_{\rm MSE}$ is the loss function minimize the mean square error between two images. $\mathcal{L}_{\rm perc}$ is the perceptual loss of two images $\hat{\textbf{l}}$ and $\textbf{l}$ defined as:
\begin{equation}
\label{eq17}
\mathcal{L}_{\rm perc}(\hat{\textbf{l}}, \textbf{l})=\mathcal{L}_1(\phi_k(\hat{\textbf{l}}), \phi_k(\textbf{l})),
\end{equation}

\noindent where $\phi_k(\cdot)$ are features extracted from $k^{\rm th}$ layer of VGG16 network \cite{simonyan2014very}. We use $k=8$ for $\mathcal{L}_{\rm perc}$.

$\mathcal{L}_{\rm SSIM}$ is a loss fucntion to maximize the structural similarity metric (SSIM) between two images \cite{snell2017learning, zhao2016loss}. SSIM for image $\hat{\textbf{l}}$ and $\textbf{l}$  is defined as: 
\begin{equation}
\label{eq18}
\rm{SSIM}(\hat{\textbf{l}}, \textbf{l}) = \frac{(2\mu_{\hat{\textbf{l}}}\mu_{\textbf{l}}+C_1)(2\sigma_{\hat{\textbf{l}}\textbf{l}}+C_2)}{(\mu_{\hat{\textbf{l}}}^2 + \mu_{\textbf{l}}^2+C_1)(\sigma_{\hat{\textbf{l}}}^2 + \sigma_{\textbf{l}}^2+C_2)},
\end{equation}

\noindent where $\mu_{\boldsymbol{\rm l}}$, $\sigma_{\boldsymbol{\rm l}}^2$, and $\sigma_{\hat{\boldsymbol{\rm l}}\boldsymbol{\rm l}}$ are the mean of $\boldsymbol{\rm l}$, the variance of $\boldsymbol{\rm l}$, and the covariance of $\boldsymbol{\rm l}$ and $\hat{\boldsymbol{\rm l}}$. $C_1$ and $C_2$ are constant values and we set them as $1e^{-4}$ and $9e^{-4}$ respectively. To evaluate SSIM, we use a sliding Gaussian window with a size equals to 11 to attain the mean $\rm{SSIM}(\hat{\textbf{l}}, \textbf{l})$:
\begin{equation}
\label{eq19}
{\rm MSSIM}(\hat{\textbf{l}}, \textbf{l}) = \frac{1}{M}\sum_{i=0}^{M}\rm{SSIM}(\hat{\textbf{l}}_i, \textbf{l}_i).
\end{equation}

The final $\mathcal{L}_{\rm SSIM}$ is defined as:
\begin{equation}
\label{eq20}
\mathcal{L}_{\rm SSIM}(\hat{\textbf{l}}, \textbf{l}) = 1 - {\rm MSSIM}(\hat{\textbf{l}}, \textbf{l}).
\end{equation}

\subsection{HDR-net: fusing multi-irradiance video into a single HDR video}
\subsubsection{HDR-net architecture}
The HDR-net, $\mathcal{D}_{rm HDR}(\cdot)$, is used to generate weight matrices $[\hat{\textbf{w}}_{\rm low}, \hat{\textbf{w}}_{\rm mid}, \hat{\textbf{w}}_{\rm high}]$ for fusing multiple LDR videos from $\mathcal{D}_{\rm LDR}$ into a single HDR video. We built $\mathcal{D}_{\rm HDR}(\cdot)$ with 5 consecutive 3D CNN residual modules ($3\times3\times3$ convolutional layers inside). The HDR-net receives 6-channel 4D tensor (6 stacked video cubes) as the input and generates 3-channel 4D tensor (3 weight cubes) as the output. 

\subsubsection{training}
$\mathcal{D}_{\rm HDR}(\cdot)$ completes the following transformation: 
\begin{equation}
\label{eq21}
\hat{\textbf{w}}_i=\mathcal{D}_{\rm HDR}(\hat{\textbf{y}}_i, \hat{\textbf{p}}_i), \quad i\in\{{\rm low, mid, high}\},
\end{equation}

\noindent where $[\hat{\textbf{p}}_{\rm low}, \hat{\textbf{p}}_{\rm mid}, \hat{\textbf{p}}_{\rm high}]$ are derived from $\mathcal{D}_{\rm LDR}(\cdot)$ outputs $[\hat{\textbf{y}}_{\rm low}, \hat{\textbf{y}}_{\rm mid}, \hat{\textbf{y}}_{\rm high}]$ using the inverse CRF function:
\begin{equation}
\label{eq22}
\hat{\textbf{p}}_i=\mathcal{R}^{-1}(\hat{\textbf{y}}_i) \quad i\in\{{\rm low, mid, high}\}.
\end{equation}.

Finally, the output weights, $[\hat{\textbf{w}}_{\rm low}, \hat{\textbf{w}}_{\rm mid}, \hat{\textbf{w}}_{\rm high}]$, are used to combine coarsely refined cubes $[\hat{\textbf{p}}_{\rm low}, \hat{\textbf{p}}_{\rm mid}, \hat{\textbf{p}}_{\rm high}]$ to form a single HDR video $\hat{\textbf{p}} \in \mathbb{R}^{N_r \times N_c \times N_t}$:
\begin{equation}
\label{eq23}
\hat{\textbf{p}} = \frac{\hat{\textbf{w}}_{\rm low} \odot \hat{\textbf{p}}_{\rm low} + \hat{\textbf{w}}_{\rm mid} \odot \hat{\textbf{p}}_{\rm mid} + \hat{\textbf{w}}_{\rm high} \odot \hat{\textbf{p}}_{\rm high}}{\sum_{i}\hat{\textbf{w}}_i}.
\end{equation}

We adopt the differentiable $\mu$-law function \cite{kalantari2019deep, chen2021hdr} tone-map the training input/output videos, before computing the loss function. To do this, we set the medium exposure $4\textbf{p}$ as the reference ground truth. Then the estimated tone-mapped video $\hat{\textbf{l}}$ and ground truth $\textbf{l}$ are:
\begin{equation}
\label{eq24}
\hat{\textbf{l}} = \frac{\rm{log}(1+\mu\hat{\textbf{p}})}{\rm{log}(1+\mu)}, \quad 
\textbf{l} = \frac{\rm{log}(1+4\mu\textbf{p})}{\rm{log}(1+\mu)},
\end{equation}

where $\mu$ is typically set as 5000. Then we minimize the loss function $\mathcal{L}(\cdot)$, described in Section V(B), to fuse HDR frames:
\begin{equation}
\label{eq25}
\mathop{\arg\min}_{\mathcal{D}_{\rm HDR}} \mathcal{L}(\hat{\textbf{l}}, \textbf{l}).
\end{equation}


\begin{figure*}[!t]
\centering
\includegraphics[width=7in]{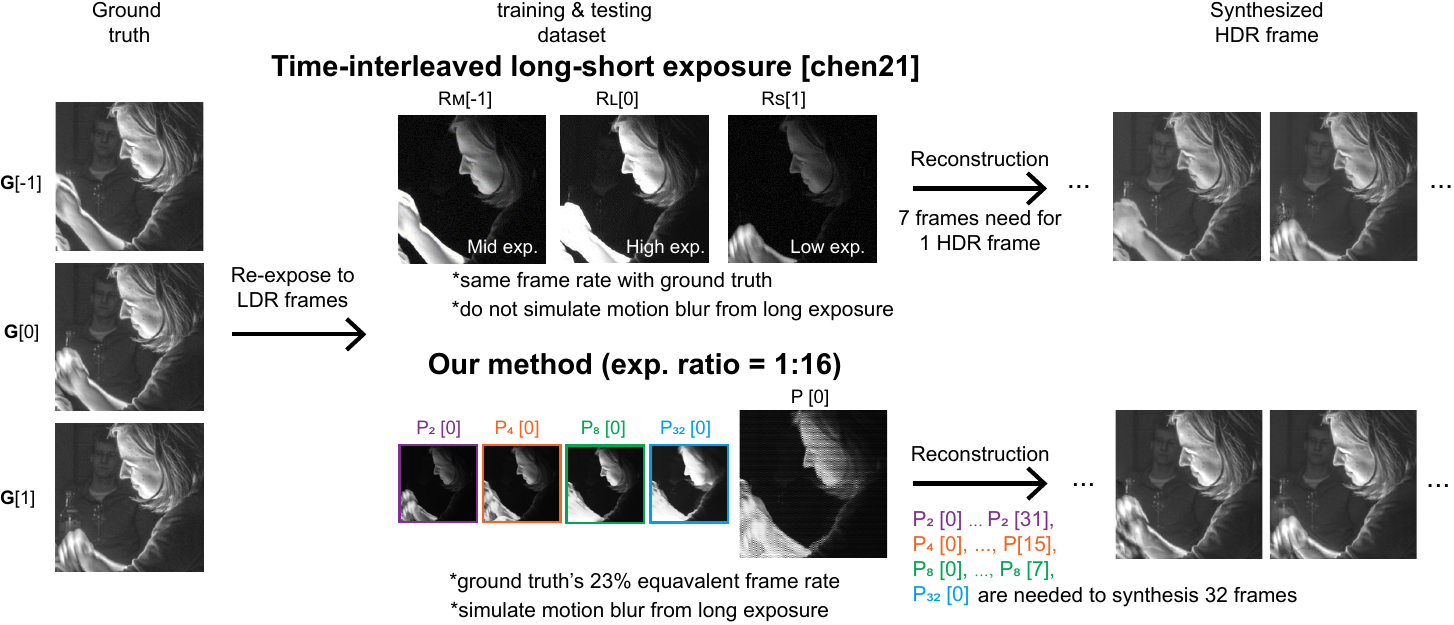}%
\caption{Comparison of Pix2HDR with the time-interleaved long-short exposure method (Chen21) at synthesis HDR videos. \textbf{A.} generation of training/testing dataset from ground truth. Chen21 transforms HDR frames into LDR frames using camera exposure functions ($\rm R_S$, $\rm R_M$, and $\rm R_L$), simulating medium, short and long exposure. This dataset does not simulate motion blur and frame rate decrease resulting from prolonged exposure. \textbf{B.} to correctly model motion blurring, we summed the ground truth HDR frames’ pixel values before transforming them through the camera response function (Rp). $\rm P_2$, $\rm P_4$, $\rm P_8$, $\rm P_{32}$, denote pixels undergoing exposure equivalent to 2, 4, 8, or 32 frames of the ground truth frames. The Pix2HDR’s dataset correctly simulates motion blur and the frame rate decrease resulting from long exposure.}
\label{fig_r3}
\end{figure*}

\section{Results}
\subsection{Comparison with previous work using HDR database}
We compared the performance of the Pix2HDR method with previous works using the hdm-hdr-2014 dataset \cite{froehlich2014creating}. We put the MPVE exposure pattern in different configurations so that the min-max pixel exposure ratios within a 2 $\times$ 2 pixel block are 1:4, 1:8, and 1:16. We then train three models to fit the measurements with new exposure ratios. To prepare the training data for the LDR-HDR networks, we first downsample the ground truth video from its resolution ($1920 \times 1080$) by half to ensure each $128\times128$ pixel patch has FOV comparable to our camera. We then crop the consecutive frame into $128\times128\times8$ (when exposure ratio is 1:4) or $128\times128\times16$ (when exposure ratio is 1:8 or 1:16) video cubes for training. The cubes have a 50\% overlap along the temporal axis. The training dataset is also augmented with 90\degree, 180\degree, and 270\degree rotated versions. 21 HDR videos are used as training data from the dataset, with one (named Poker Fullshot) used for testing. LDR-net and HDR-net training were completed on a Nvidia RTX3080, and the average inference time to generate one frame is 2.5ms, which supports 400Hz real-time HDR video streaming. The code and real-world video demonstrations can be found at: https://github.com/csmslab/Pix2hdr.

We compared our Pix2HDR method of different min-max pixel exposures with the interleaved long-short exposure sampling \& deep learning-based synthesis method demonstrated by Chen et. al. (“Chen21” \cite{chen2021hdr}). To simulate the long and short exposure frames, Chen et. al. transformed HDR ground truth videos back into LDR videos at interleaved short, medium, and long exposures, utilizing nonlinear camera response functions (denoted here as $\rm R_S$, $\rm R_M$, and $\rm R_L$, \textbf{Fig. \ref{fig_r3}}). To simulate overexposure, they applied a thresholding technique that clamps pixel values at predetermined levels. These generated LDR images then served as inputs for a neural network designed to reconstruct HDR frames. This LDR video generation process mimics the frames at different exposures. However, it notably fails to replicate one critical aspect of long exposure: motion blur. As indicated in \textbf{Fig. \ref{fig_r3}} ground truth frame G[1] has the same motion blur as the generated LDR input $\rm R_S[1]$. As a result, the trained neural network does not have to resolve motion blurring caused by prolonged exposure. While this could yield high-performance relative to the ground truth in the test setup, it might hurt the performance when the method is applied to real imaging applications. 

To correctly account for motion blurring in generating input data in our MPVE sampling, we summed the ground truth HDR frames’ pixel values before transforming them through the camera response function. To illustrate, in MPVE with an exposure ratio of 1:16, $\rm P_2$, $\rm P_4$, $\rm P_8$, $\rm P_{32}$ denote pixels undergoing exposure equivalent to 2, 4, 8, or 32 frames of the ground truth frames (\textbf{Fig. \ref{fig_r3}}) Overexposure is also simulated by thresholding to clamp pixel values at predetermined levels. The level is chosen to match that of Chen21 to ensure a similar amount of overexposure. As a result, long exposure pixels, such as $\rm P_8$, $\rm P_{32}$ are corrupted by motion blur and overexposures.

It is also worth noting that the pixel sampling rate from MPVE sampling is much lower than that of the ground truth. This is compared to Chen21, where the frame rate of the generated LDR frames is the same as the HDR ground truth. The inclusion of motion blur, combined with reduced frame rate, increases the task complexity of our method compared to that of Chen21. We compared the PSNR and SSIM of the synthesized HDR video to evaluate the HDR synthesis performance from both methods. We measured two additional metrics HDR-VDP \cite{mantiuk2011hdr} and HDR-VQM \cite{narwaria2015hdr}. HDR-VDP evaluates image pair differences from arbitrary luminance ranges, and HDR-VQM measures the same differences for the entire video sequences. The higher the number, the closer the synthesized frames are to the ground truth.

\begin{figure*}[!t]
\centering
\includegraphics[width=7in]{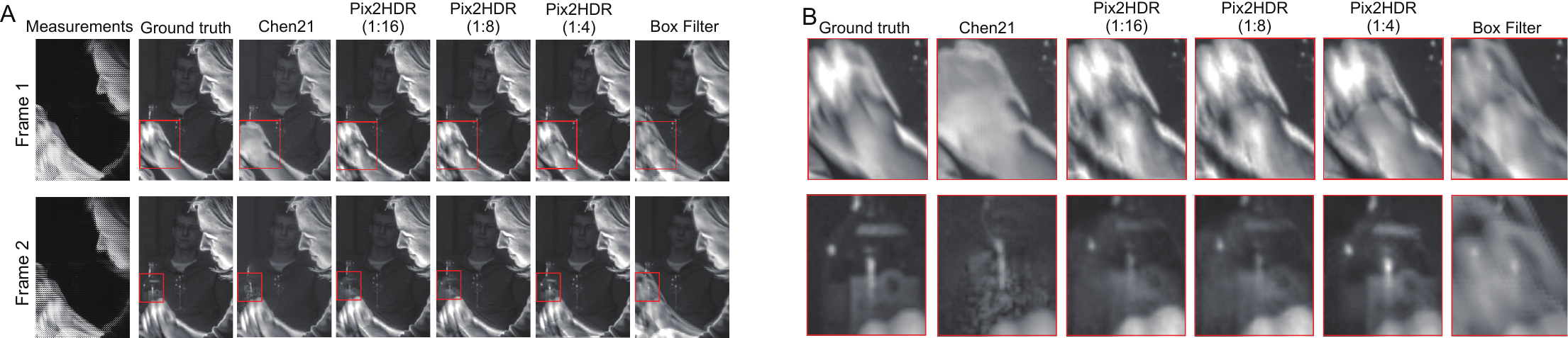}%
\caption{Example of Pix2HDR and existing methods at resolving HDR videos. \textbf{A.} Full frame comparison, \textbf{B.} Zoomed in at the motion. Interleaved long-short exposure sampling can result in large spatiotemporal information loss. In contrast, MPVE sampling can maintain optimal pixel exposure in all the local video regions and time steps, thereby preventing loss of spatiotemporal information due to overexposure, low SNR, or blurring. Besides, prolonging exposure duration on pixels introduce more artifact.}
\label{fig_8}
\end{figure*}

The results show that our Pix2HDR method at different MPVE configurations consistently outperforms the interleaved long-short exposure method (\textbf{Table \ref{tab1}}). This superiority is attributed to the MPVE sampling’s capacity to maintain optimal pixel exposure in all the local video regions and time steps, thereby preventing loss of spatiotemporal information due to overexposure, low SNR, or blurring. This is evident in the test image, P[0], which consists of pixels at exposure 2, 4, 8, and 32 (\textbf{Fig. \ref{fig_r3}}). The preservation of spatiotemporal information during sampling greatly eases the synthesis neural network’s effort to recover them during HDR video generation (\textbf{Fig. \ref{fig_8}}).

In contrast, interleaved long-short exposure sampling can result in large spatiotemporal information loss. As illustrated in the test images, in $\rm R_L[0]$ (\textbf{Fig. \ref{fig_r3}}), long pixel exposure completely saturates the details of the hand, while in $\rm R_S[1]$, the bottle in the background is sampled with low SNR due to short exposure. Although Chen21’s neural networks utilize 3 frames before and after the target frame to recover the lost information, this approach often results in suboptimal outcomes due to a significant loss of information at the sampled frames (\textbf{Fig. \ref{fig_8}}).

We also synthesized HDR videos from MPVE sampling outputs using a spatial box filter \cite{nayar2000high}, yielding inferior performance compared to the results from the HDR-LDR network. In the example, it is unable to separate the background bottle and the hand. This demonstrates the superior performance of the Pix2HDR method is attributed to both the MPVE sampling and the HDR-LDR video synthesis network.

\begin{table*}
\begin{center}
\caption{Comparative metrics of Pix2HDR with existing methods at resolving HDR videos}
\label{tab1}
\begin{tabular}{| c | c | c | c | c | c | c |}
\hline
  Name (exp. ratio)& Sampling method & Synthesis method & PSNR(dB) & 
SSIM & HDR-VDP & HDR-VQM\\
\hline
Box filter \cite{nayar2000high} (1:4) & Spatially varying exposure & Spatial box filter & 32.99 & 0.9679 & 95.38 & 50.86\\
\hline
Chen21 \cite{chen2021hdr} (1:16) & Long-short exposure & 2D CNN + Flow net & 40.20 & 0.9364 & 95.03 & 54.61\\
\hline
Pix2HDR (1:16) & MPVE & 3D CNN & 42.66 & 0.9782 & 95.54 & 66.95 \\
\hline 
Pix2HDR (1:8) & MPVE & 3D CNN & 42.94 & 0.9786 & 95.55 & 67.62 \\
\hline 
Pix2HDR (1:4) & MPVE & 3D CNN & \textbf{43.51} & \textbf{0.9821} & \textbf{95.60} & \textbf{68.96} \\
\hline 
\end{tabular}
\end{center}
\end{table*}

\begin{figure}[!t]
\centering
\includegraphics[width=3.49in]{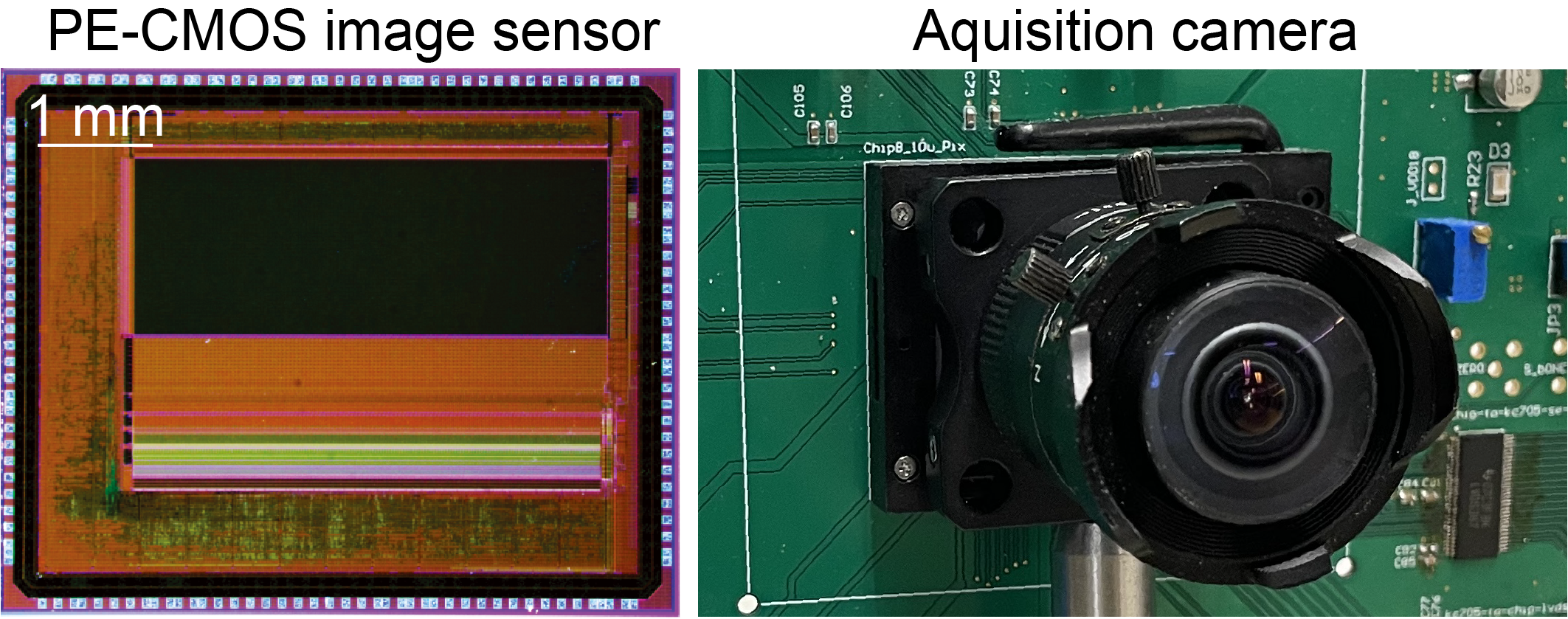}
\caption{PE-CMOS image sensor and the acquisition camera}
\label{fig_9}
\end{figure}

\begin{figure*}[!t]
\centering
\includegraphics[width=7in]{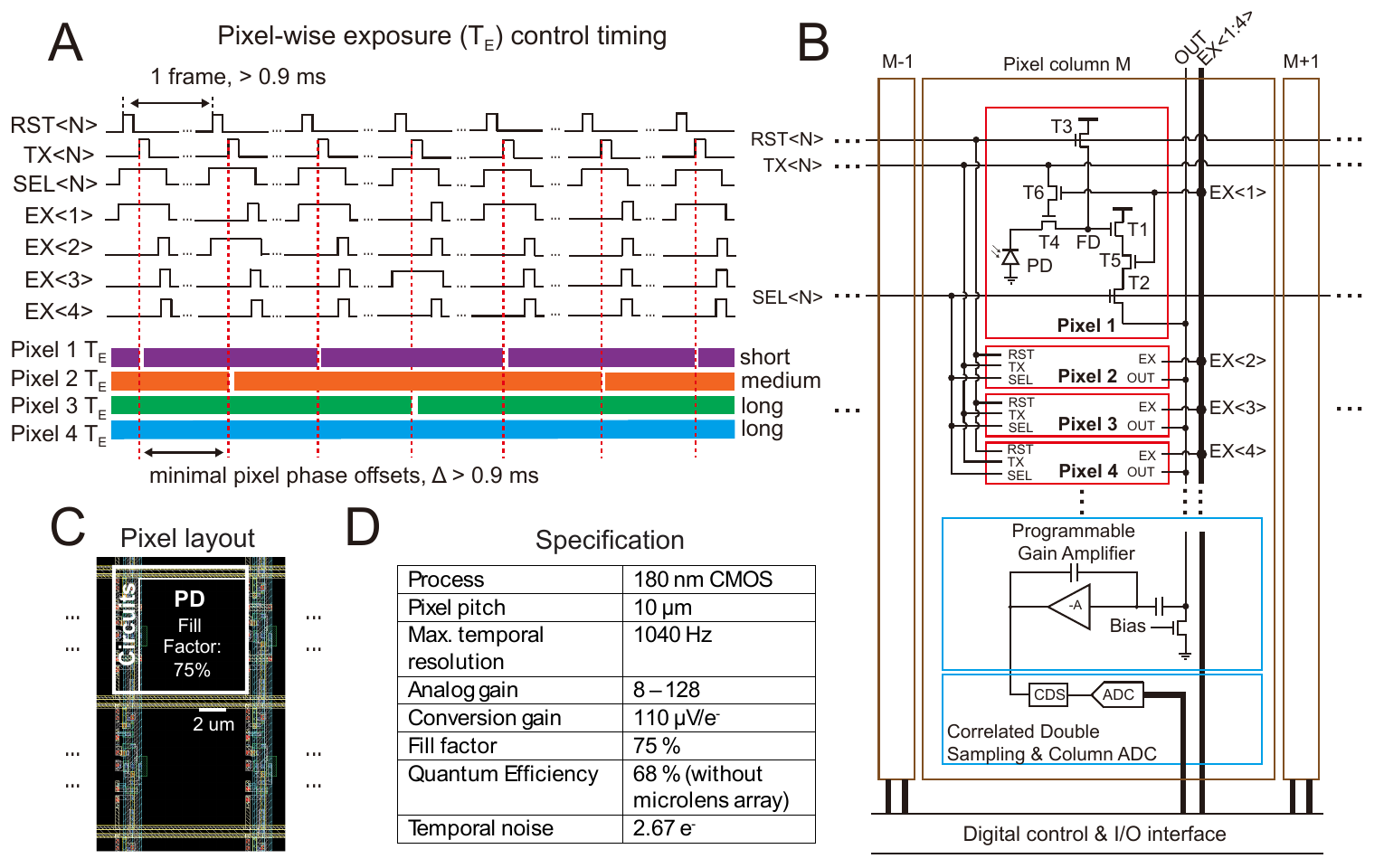}
\caption{PE-CMOS design and control timing. \textbf{A.} an example timing to put pixel 1 – 4 at different exposure duration and phase. \textbf{B.} the PE-CMOS schematic. Pixels are organized in groups consisting of 4 row and 1 column that shares row control signals (RST, TX, and SEL). Each column shares one amplifier and ADC. The pixel design features 6 transistors (T1 – T6) to provide pixel-wise exposure control while maximizing the pixel fill factor. \textbf{C.} Pixel layout. \textbf{D.} sensor specifications table. }
\label{fig_R5}
\end{figure*}

The Pix2HDR’s LDR-HDR net’s superior synthesis performance is also evident when it is compared with the spatial filtering \cite{nayar2000high}. Both methods acquired pixel-wise outputs using MPVE sampling but synthesized HDR videos with different techniques. The spatial box filter processes each frame independently without considering its temporal correlations with adjacent frames. The reconstructed video sequence thus suffers from blurring from long exposure pixels (\textbf{Fig. \ref{fig_8}}). In addition, each frame has a lower resolution due to the low-pass nature of the filter. In contrast, Pix2HDR utilizes LDR-HDR networks to covert pixel-wise outputs directly to high-resolution HDR video cubes, using learned weights that minimize blurring, resulting in notably enhanced performance in all the metrics (\textbf{Table \ref{tab1}}).

\begin{figure}[!t]
\centering
\includegraphics[width=3.45in]{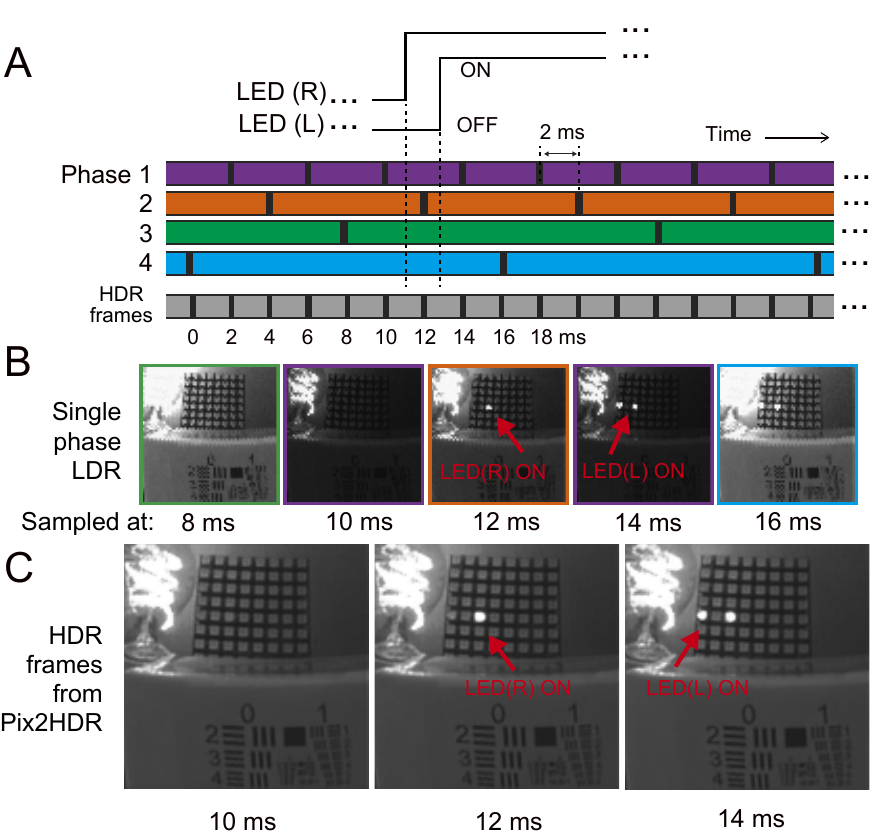}
\caption{Capturing transient events faster than individual pixels’ speed in an HDR setting. \textbf{A.} Experiment setup: two LEDs (R and L) are switched on sequentially with $\sim$2 ms delay. PE-CMOS pixels are configured with 4 ms, 8 ms and 16 ms exposure time. Pixel-wise phase-offset is set to 2 ms and HDR frames are also reconstructed at 2 ms temporal resolution. \textbf{B.} Single phase LDR frames. Each individual pixel’s speed ($\geq$ 4 ms sampling period) is insufficient to capture events that are spaced $\sim$2 ms in time. But this high-speed event is captured by the phase-offset pixels, achieving temporal resolution of 2 ms. \textbf{C.} HDR frames by Pix2HDR portraits the high-speed event with 2 ms temporal resolution.}
\label{fig_10}
\end{figure}

\subsection{Pix2HDR performance at capturing HDR scenes}
Next, we assess the Pix2HDR's performance at capturing real-world HDR scenes. Specifically, we evaluate its temporal resolutions, dynamic range, motion deblurring, and ability to detect object motion under low-light conditions and against bright backgrounds — essential metrics for high-speed HDR imaging.

We built a camera using the PE-CMOS image sensor that implements the MPVE pixel-wise sampling pattern (\textbf{Fig. \ref{fig_9}}). The PE-CMOS sensor is described in detail in our previous publication \cite{zhang2023pixel}, here, we summarize the readout control of the multi-phase sampling scheme. The PE-CMOS pixel design (\textbf{Fig. \ref{fig_R5}B}) comprises 6 transistors (T1 - T6) and one pinned photodiode (PD). T1 to T4 form the standard rolling shutter 4T-pixel design. During pixel exposure, the PD converts incoming photons into electrons. When the PD is ready to be sampled, T3, controlled by a row reset (RST) signal, first resets the voltage on the floating diffusion (FD) node. T4 (controlled by TX) then moves the electrons from the PD to FD, producing a change in voltage. The voltage signal is buffered by T1 and connected to the readout circuits by T2 (controlled by SEL), where the column readout circuitry samples it.

We incorporated two additional transistors (T5 and T6) into the PE-CMOS pixel design to enable pixel-level exposure programmability. T5 and T6 operate as switches with input driven by column signals, EX. In PE-CMOS, each set of row signals (TX, RST, SEL) selects 4 rows of pixels as candidates for readout. At each column, out of these K pixels, only the pixel with both T5 and T6 activated will end its exposure phase and be sampled by the column circuitry. Meanwhile, the other pixels, with their T5 and T6 remaining off, will continue their exposure. T5 and T6 are controlled by column bus lines EX$\langle1:4\rangle$ placed at each column. Pixel-wise exposure control can be achieved by synchronizing the EX signals with row signals (RST, SEL, and TX). An example pixel-wise operation diagram is shown in \textbf{Fig. \ref{fig_R5}A}. For a group of K pixels, the pixel ends its exposure (at the time marked by dotted red line in \textbf{Fig. \ref{fig_R5}A}) whenever its corresponding EX signal is high during the readout operation driven by the signals: RST$\langle \rm N\rangle$, TX$\langle \rm N\rangle$ and SEL$\langle \rm N\rangle signal$.

The PE-CMOS pixel is implemented using a commercial CMOS image sensor process. The design can be easily incorporated into any CMOS image sensors design to achieve flexible pixel-wise control. The pixel is optimized for highspeed low-light applications with high PD fill factor (75\%), high conversion gain (110 $\rm \mu V/e^-$), and low noise (2.67 $\rm e^-$). Its layout and specification are shown in \textbf{Fig. \ref{fig_R5}C} and \textbf{D}. The PE-CMOS sensor interfaces with an FPGA (Xilinx Kintex-7), which controls the signal timings and receives the pixel outputs from the chip and then transfers them to a host PC through a high-speed PCIe bus. We implement the acquisition software on the PC using the Open-Ephys ONIX API \cite{newman2023unified}, designed for high-speed PCIe interfaces.

\subsubsection{Temporal Resolution}
We first evaluate the Pix2HDR’s temporal resolution at sampling high-speed events. We direct the camera at an LED array, where sequential activation of two LEDs (R and L) takes place with a short time delay (\textbf{Fig. \ref{fig_10}A}). Our objective is to demonstrate Pix2HDR’s capability to detect the transient ‘ON’ events with time resolution surpassing the camera's shortest exposure time. 

We set up the MPVE with pixel exposures of 4 ms (short), 8 ms (medium), and 16 ms (long). These pixels are organized into four phases, with a minimal phase offset of 2 ms between adjacent pixels (\textbf{Fig. \ref{fig_10}A}). The LED activation time delay is configured to $\sim$2 ms — much faster than the shortest pixel exposure time. 

As expected, even the pixels with the shortest exposure time fail to discern the individual LED’s activation events: pixels at phase 1, sampled at 10 ms and 14 ms, do not capture LED R's activation. This event, however, is unambiguously isolated by pixels with phase offsets at phase 2, sampled at 12 ms (\textbf{Fig. \ref{fig_10}B}). Between phase-shifted pixels, Pix2HDR can resolve a temporal resolution of 2 ms (500 Hz), two times faster than the shorted pixels exposure at 4 ms (250 Hz). Consequently, Pix2HDR can synthesize HDR video sequences at a time resolution of 2 ms with 500 FPS (\textbf{Fig. \ref{fig_10}C}).

\begin{figure*}[!t]
\centering
\includegraphics[width=7in]{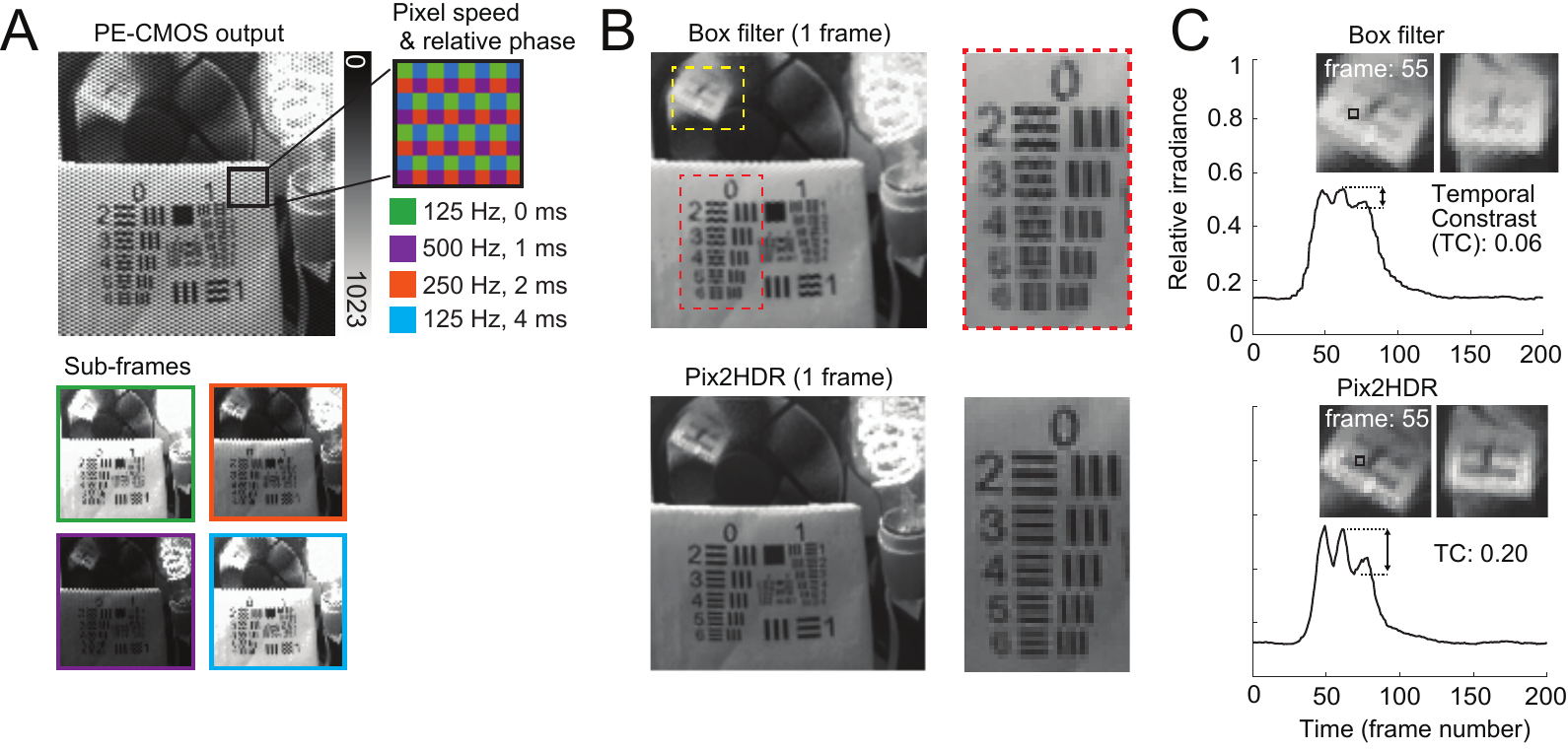}
\caption{Capturing HDR video with Pix2HDR. \textbf{A.} Direct measurement from PE-CMOS. In any $2\times2$ pixel patch, pixels are sampling at 2 - 8 ms exposure with 125 - 500 Hz sampling rate, and relative phase shifts of 0 - 4 ms. \textbf{B.} Recovered HDR frames (using box filter and Pix2HDR method). The frame from our Pix2HDR has higher spatial contrast and lower blurring to resolve to moving letter “H” with sharper resolution. \textbf{C.} Temporal resolution measurements. We fix a single pixel and record its irradiance value across frames. Our Pix2HDR method shows larger irradiance variation corresponding to the motion of the “H” and thereby results in a much higher temporal contrast (TC = 0.2) than box filter (TC = 0.06).}
\label{fig_11}
\end{figure*}

\subsubsection{Dynamic range and deblurring performance}
Next, we assess the Pix2HDR’s performance at capturing motion in an HDR context. We set up a scene with diverse intensity and motions: a bright area created by a light bulb, a dark region with a USAF target, and a fan rotating at a speed of $\sim$300 rpm, with a letter “H” printed on it. To capture this scene, we configured the MPVE with pixel exposures of 2 ms (short), 4 ms (medium), and 8 ms (long), with a 1 ms minimal pixel-wise phase offset (\textbf{Fig. \ref{fig_11}A}). Pixels raw outputs are displayed by grouping into subframe with 1/4 spatial resolution. Short exposure pixels capture blurred free motion and avoid overexposure at the bright region at the cost of lower SNR. Pixels with longer exposures enhance the SNR at the expense of blurring and overexposure. 

We combined the pixel-wise outputs to form an HDR video sequence with the LDR-HDR nets and then compared the synthesized sequences with those constructed through a spatial box filter. The Pix2HDR results have better HDR spatial contrast and resolution, notably in the reconstructed light blub and the USAF target (\textbf{Fig. \ref{fig_11}B}). To quantify the motion trajectory, we take the intensity of a single pixel located along the path of the rotation motion. As the letter “H” passes through it, it generates a time-varying intensity, $v$, corresponding to the letter’s strokes’ trajectory (\textbf{Fig. \ref{fig_11}C}). We  slightly modified the Michelson contrast \cite{peli1990contrast} to temporal axis and calculated the temporal contrast (TC) of $v$, during a pass-through event:
\begin{equation}
\label{eq26}
\text{TC}=\frac{v_{\rm peak}-v_{\rm through}}{v_{\rm peak}+v_{\rm through}}.
\end{equation}

The TC quantifies the overall detectability of a motion pattern. It is affected by a few spatial and temporal factors: SNR, spatial contrast, and motion blurring, as smoothed edges can lead to TC value degradation. Pix2HDR's synthesized HDR frames exhibit significantly improved TC compared to box-filtered outputs, clearly resolving the edges of the fast-moving letter "H" (\textbf{Fig. \ref{fig_11}C}).

\begin{figure*}[!t]
\centering
\includegraphics[width=6.5in]{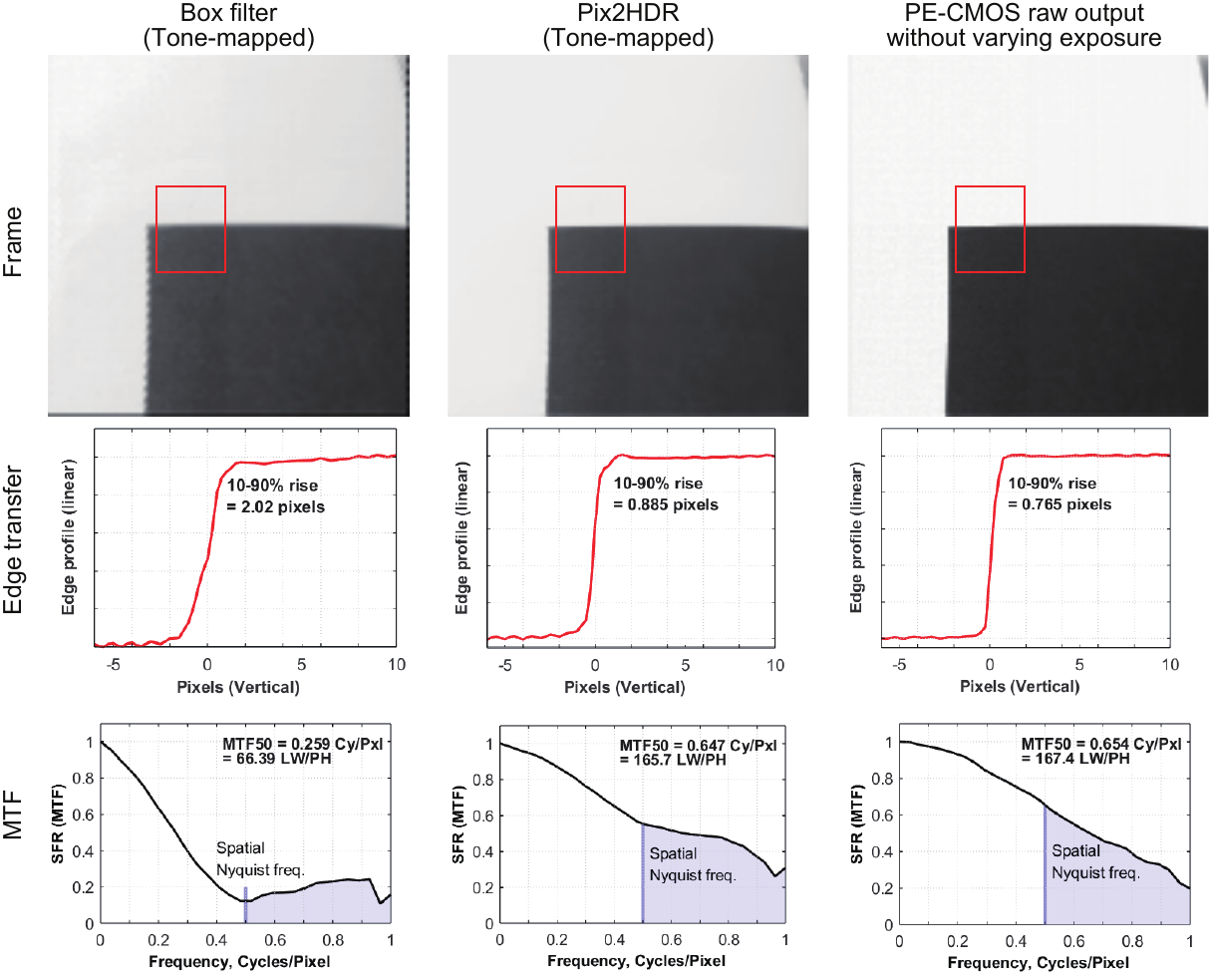}%
\caption{Comparison of edge transfers and MTFs among HDR frames (from box filter and Pix2HDR) and direct LDR output from PE-CMOS. To qualify the spatial resolution, we measure the edge transfers and MTFs in a region with abrupt transition from dark to bright. Compared with the raw output from PE-CMOS, the frame from our Pix2HDR only has tiny distance prolongation in 10-90\% rise rate and little drop in MFT50 value, which means our Pix2HDR method only loses neglectable spatial resolution after reconstruction. Meanwhile, the box filter gives much worse scores in the metrics and even generates jagged vertical edge in the recovered frame.}
\label{fig_12}
\end{figure*}

\subsubsection{Spatial resolution}
To quantify Pix2HDR’s spatial resolution, we measured its modulation transfer functions (MTFs) using the slanted-edge method \cite{estribeau2004fast}. The target scene contains a single edge that abruptly transitions from dark to bright. We then synthesized the HDR frames from MPVE pixel-wise outputs with the Pix2HDR and box filtering. 

The edge transfer and MTF all indicate higher spatial resolution for Pix2HDR at preserving sharp edges (\textbf{Fig. \ref{fig_12}}). MTF50 denotes the spatial frequency where contrast drops to 50\% of its original value — a higher MTF50 score signifies a sharper, more detailed image. As the recovery strategies (box filter and Pix2HDR) inevitably cause sharpness loss in above frames, we also acquired the full-resolution frame using PE-CMOS with fixed pixel-wise exposures and set its scores as the theoretical limits to the camera optics and pixel pitch. Pix2HDR achieves superior edge transfer (0.885 pixels for 10-90\% rise rate) and MTF50 values (0.647 cycles/pixel), surpassing those from the box-filtered output and approaching theoretical limits of 0.765 pixels and 0.654 cycles/pixel, respectively.

\begin{figure*}[!t]
\centering
\includegraphics[width=7in]{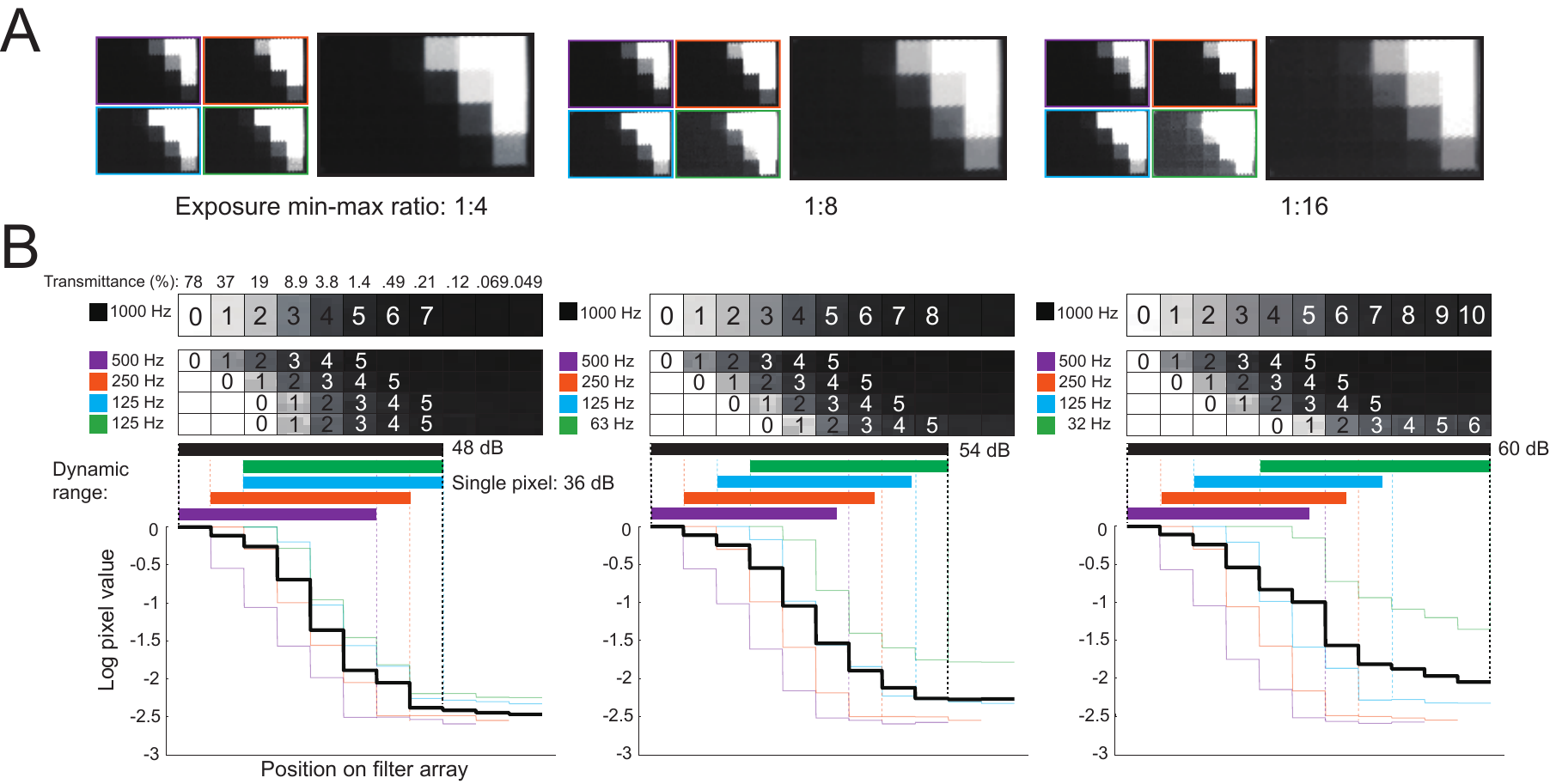}%
\caption{Video dyanmic range enhancement through MPVE configurations. \textbf{A.} MPVE sampling of an ISO-21550 dynamic range film target. \textbf{B.} Center areas of rectangles on the targets are truncated and lined up for comparison. Pixels at 500, 250, 125, 63 and 32 Hz are configured into MPVE sampling with pixel min-max ratio of 1:4, 1:8 and 1:16, achieving 48 – 60 dB of dynamic range, 12 – 24 dB higher than a single pixel’s dynamic range (36 dB).}
\label{fig_r2}
\end{figure*}

\subsubsection{Dynamic range enhancement}
To quantify the dynamic range enhancement from our method, we set the PE-CMOS pixels in 3 different configurations, with min-max pixel exposure ratios of 1:4, 1:8, and 1:16 using MPVE sampling pattern with pixels at 500, 250, 125, 63, and 32 Hz (\textbf{Fig.\ref{fig_r2}}). We then reconstructed HDR videos at 1000 Hz using our LDR-HDR network. We first sampled a fixed target (the ISO-21550 dynamic range film target) of different transmittances to evaluate the dynamic range performance of these configurations (\textbf{Fig.\ref{fig_r2}A}). The pixel’s dynamic range is determined by the difference between its maximum and minimum resolvable transmittance values, indicated by the horizontal color bar in the figures. The PE-CMOS pixels are designed for high-speed (1000 Hz) and high-sensitivity imaging applications with low full-well capacity, high converstion gain and analog amplifications, which limits each pixel’s dynamic range to ~36 dB. With MPVE, the combined video at 1000 Hz improves the dynamic range by 12 – 24 dB, depending on the pixel configurations (\textbf{Fig.\ref{fig_r2}B}). The higher the min-max pixel exposure ratio, the higher better the dynamic range improvement. The pixel exposure ratio could be increased further, as PE-CMOS has no limitation in extending a pixel’s exposure indefinitely. But a consideration that would discourage excessively long-pixel exposure is the high-speed performance, as prolonged exposure will result in extra motion blur that could affect the performance of the synthesized HDR video. Some examples at different MPVE configurations are shown in supplemental videos.

\begin{figure*}[!t]
\centering
\includegraphics[width=7.2in]{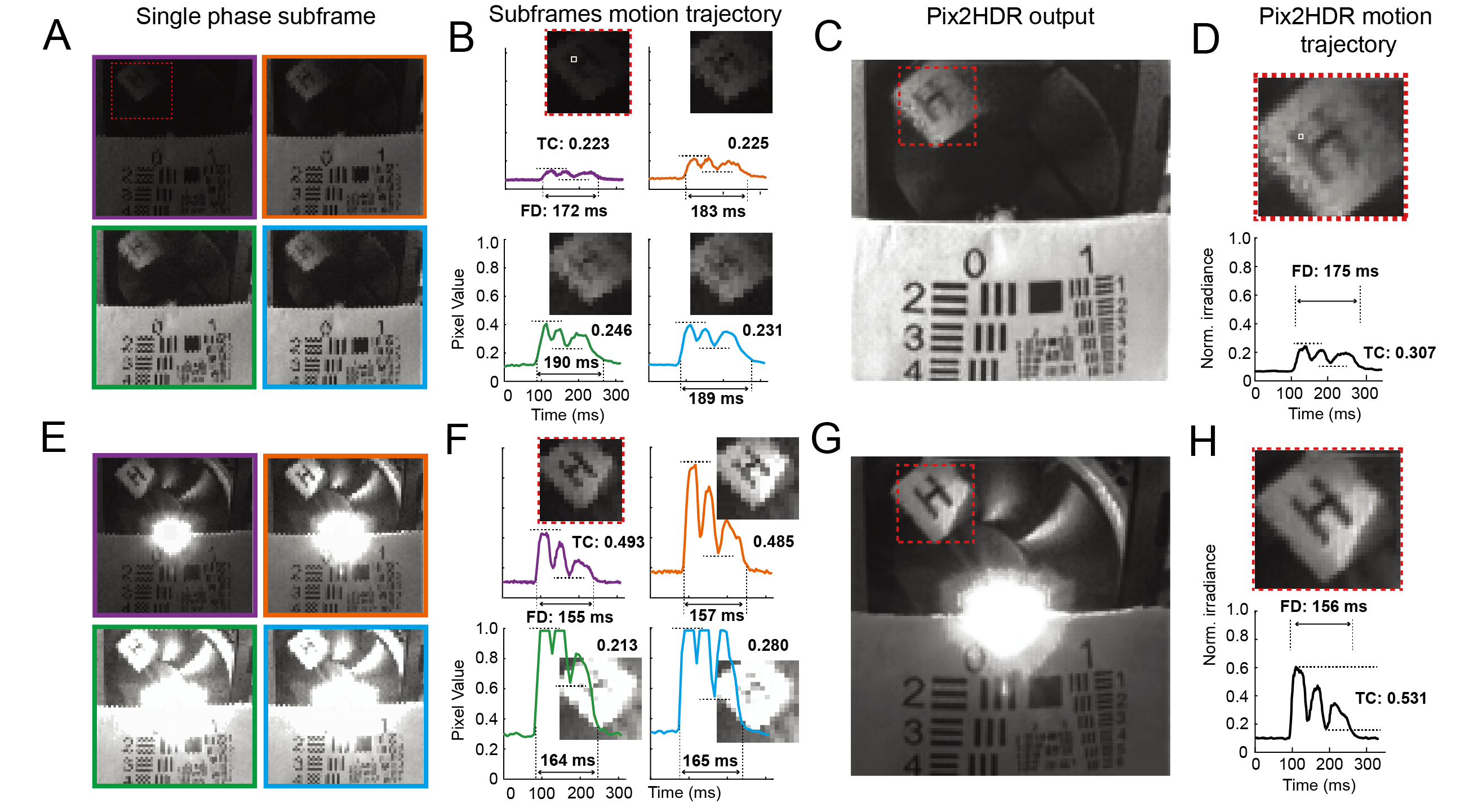}%
\caption{Imaging fast motion in dark and against bright background. \textbf{A.} Subframes of PE-CMOS outputs under low illumination conditions. \textbf{B.} Time series of at a single pixel value indicating the advantage of different exposures at capturing motion: Long exposure subframes have higher temporal contrast (TC) at resolving the motion of the letter “H”, while short exposure subframes have the shortest full duration (FD), indicating minimal blurring. \textbf{C-D.} Pix2HDR output (\textbf{C}, single frame, and \textbf{D}, pixel time-series). Pix2HDR combines the strength of long and short exposures, resulting in the highest TC compared. Its FD is close to that of short exposure subframes, indicating superior performance at resolving motion in the dark. \textbf{E-H.} Pix2HDR at resolving motion against the bright background. Its high TC and low FD indicates  Pix2HDR’s robust performance in resolving motion amid overexposures.}
\label{fig_13}
\end{figure*}

\subsubsection{Motion detection under low-light conditions}
We then showcase that Pix2HDR can harness the benefits of short and long pixel exposures to track movement, particularly for detecting fast motion in low-light conditions - a challenging task for video acquisition. We dim the ambient light for the demonstration and set a fan with the “H” logo printed on it into a rotational motion (\textbf{Fig. \ref{fig_13}A-D}). This creates a low-light condition for resolving motion in the dark. We sampled the scene with the MPVE using exposures of 2 ms (short), 4 ms (medium), and 8 ms (long), with a 1 ms minimal pixel-wise phase offset. 

For metrics, in addition to TC, we also calculate the motion trajectory's full duration (FD) to quantify blurring. We define FD as the timespan between 15\% of the peak irradiance measured from the rising and falling edges (\textbf{Fig. \ref{fig_13}B}). The shorter the FD, the less the blurring in the motion trajectory. 

In low ambient light conditions, pixels at different exposures have distinct advantages in terms of temporal contrast or motion blurring. Long exposures pixels show the highest temporal contrast (TC: 0.246) in comparison to those undergoing medium and short exposures (0.223) (\textbf{Fig. \ref{fig_13}B}). Meanwhile, short exposures have the minimum amount of motion blurring (FD: 172 ms), compared with medium and long exposures at 183 - 190 ms. Both the TC and FD are pure temporal metrics derived from a single pixel. They effectively resemble the outcome if cameras with the same exposures were employed. 

From the subframes, the Pix2HDR’s LDR-HDR net synthesizes a high-resolution video from pixel-wise outputs, resulting in enhanced temporal contrast (TC: 0.307) while simultaneously maintaining minimal blurring (FD: 175 ms). This highlights the Pix2HDR's advantage over single-exposure cameras in tracking motion trajectories in low light conditions. 

\subsubsection{Motion detection with bright background}
Last, we demonstrate the PixHDR’s ability to resolve motion against a bright background to evaluate its ability to capture fast movement amid overexposures. Using the same setup, we turned on a bright LED light in the middle of the FOV (\textbf{Fig.\ref{fig_13} E-H}). The bright light produces various degrees of blooming effects at each subframes.

While the additional light improved the temporal contrast of short and medium exposure (TC: 0.493 and 0.485), the blooming effects caused large overexposures in subframes with long exposures. The clipping of pixel values leads to long-exposure subframes having low temporal contrast (TC: 0.213 and 0.280). Pix2HDR’s synthesized HDR videos are robust against overexposures. The temporal contrast value is even enhanced (TC: 0.531) while the full duration (FD: 156 ms) is kept short, indicating minimal motion blurring. 

These benchtop demos in this section seek to mimic challenging conditions encountered during real-world vision applications, such as motion capture at night and against a bright headlight of incoming vehicles. The results of the demonstration showed the advantages of Pix2HDR at capturing motion during extreme conditions of low light and against bright background, having better motion detection performance than conventional cameras with fixed pixel exposures.

\section{Discussion}

\begin{figure}[!t]
\centering
\includegraphics[width=3.45in]{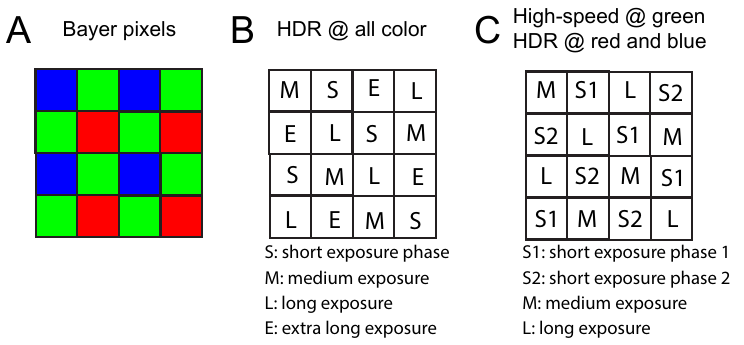}%
\caption{Adapting MPVE sampling to color videos. \textbf{A.} Commonly used Bayer filters for color image sensors. \textbf{B.} A possible MPVE arrangement to achieve HDR video at all color channels. \textbf{C.} A possible MPVE arrangement to achieve highspeed sampling at one color channel, and HDR at other color channels.}
\label{fig_r1}
\end{figure}

We have demonstrated that by combining a versatile pixel-wise sampling method with the strength of the neural networks in decoding complex light and motion patterns, Pix2HDR significantly outperforms existing methods in capturing motion in HDR settings. This makes Pix2HDR as an ideal solution for motion-tracking applications in robotics, AR/VR, and autonomous vehicles, among other vision applications. Our sensor is also compact and power-efficient. The pixel-wise sampling is implemented using the pixel circuitry and it achieves high temporal resolution using low data rate. These features make Pix2HDR applicable for mobile applications. Several aspects of Pix2HDR can be further improved: 

\textit{Adaptation to color videos}: The proposed MPVE sampling method and the LDR-HDR network are also adaptable for color video through the integration of RGB Bayer filter patterns (commonly used for color sensor) onto the PE-CMOS image sensor’s pixel array (\textbf{Fig.\ref{fig_r1}A}). To achieve high dynamic range (HDR) across all color channels, we can configure pixel exposures at short, medium, long, and extra-long durations for RGB color pixels respectively (\textbf{Fig.\ref{fig_r1}B}). The LDR-HDR synthesis network could then be trained separately for each color channel or collectively for the combined RGB. The capability to adjust exposure at the pixel level also allows each color channel to operate at different speeds, providing multi-temporal resolution suited for multi-spectral applications. For instance, we can configure the green pixels to sample at high speed with short exposures in two phases, while maintaining HDR in the blue and red channels with long exposures (\textbf{Fig.\ref{fig_r1}C}). This approach is particularly beneficial for fluorescence imaging, where it can be tailored to accommodate fluorescent indicators with distinct spectral characteristics, time constants, and dynamic ranges.

\textit{LDR-HDR network improvements}: We also plan to improve the performance of our synthesis method in future work. Currently, our LDR-HDR network features a modular architecture: The HDR network is for HDR synthesis, and the LDR network undertakes various tasks, including spatiotemporal up-sampling, deblurring, and rendering LDR videos at low, medium, and high exposure levels. We aim to refine our approach to each task separately before merging them into multiple LDR frames for input to the HDR network. For instance, we could optimize the network to deblur and temporally upsample the long-exposure pixels, similar to the methods described in \cite{jin2018learning, purohit2019bringing}. However, unlike \cite{jin2018learning, purohit2019bringing}, which depends on a neural network to upsample and reconstruct blurred frames — a challenging inverse problem — our method may leverage MPVE sampling of short-exposure pixels to provide high temporal resolution, enhancing motion estimation and deblurring. Similarly, other deep learning based architectures could be developed to recover overexposure and improve the SNR of the LDR frames, prior to video synthesis using the HDR network. 

On the synthesis side, the training steps can be made more generalizable. Currently, when creating the training dataset, camera-specific CRFs are used to map ground truth videos to each camera's output. This specificity ties the trained network weights to individual cameras. A more generalized approach could involve training the network with an overall gamma correction and mapping the captured frames from each camera to a pre-defined transfer function before network migration. 

\textit{Encoding of MPVE outputs and synthesized HDR videos}: In the MPVE sampling method, neighboring pixels operate at different frame rates and exposures, resulting in large pixel intensity differences at the small region of the video frames that might not be efficient for the standard encoding method (e.g. \textbf{Fig. \ref{fig_11}A}). To efficiently encode MPVE outputs, pixels at the same exposure (subframes) should be encoded together. This will produce four encoded video streams, each with 1/4 of the original spatial resolution. In our future work, we plan to investigate how the compression ratio of these encoded streams impacts the quality of the synthesized High Dynamic Range (HDR) videos.

The HDR video output from the LDR-HDR network can be encoded using standard video encoding formats for further processing, transmission, and storage. The choice of encoding format depends on the bit-depth of the videos, which in turn is dictated by their dynamic range. For example, the PE-CMOS sensor with MPVE configuration of min-max pixel exposure ratio of 1:16 has a dynamic range of 60 dB (\textbf{Fig. \ref{fig_r2}}). This range corresponds to a 10-bit digital format, which can be efficiently encoded using the HEVC formats. Depending on the pixel characteristics, the same MPVE configuration may have lower or higher dynamic ranges, resulting in different bit depth, and determine its encoding schemes.

\indent In summary, the Pix2HDR approach aligns with the general trend of image sensor technology development. With advancements in CMOS technology, which enable the reduction of transistor size and allow 3D integration of the pixel array and processing chips, the next generation of image sensor development is emphasizing pixel-level reconfigurability with edge processing. By combining versatile pixel-wise sampling with learning-based HDR synthesis methods that can be implemented at the edge, Pix2HDR enhances the adaptability of vision systems in dynamic conditions, addressing the growing demands of vision applications.


\bibliographystyle{IEEEtran}
\bibliography{IEEEabrv, egbib}

\vspace{-33pt}

\begin{IEEEbiography}[{\includegraphics[width=1in,height=1.25in,clip,keepaspectratio]{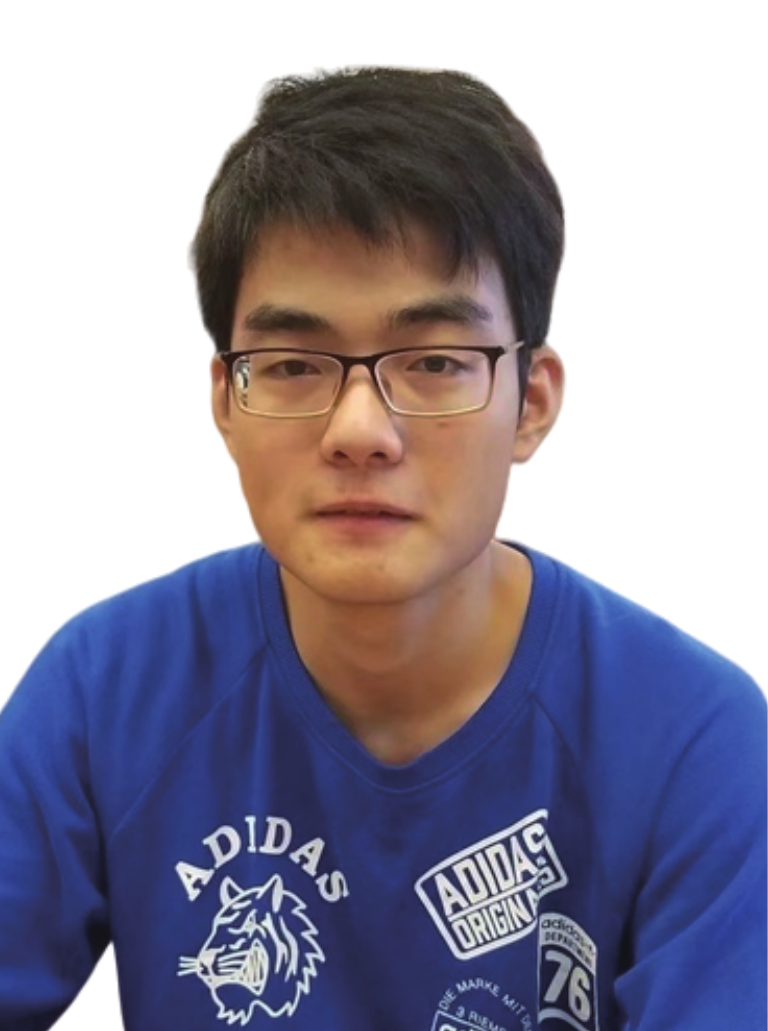}}]{Caixin Wang}
received his B.Eng. in Electrical and Electronic Engineering from University of Nottingham in 2020, and his M.S.E in Electrical and Computer Engineering from Johns Hopkins University (JHU) in 2022. He is currently working toward the Ph.D degree with Johns Hopkins University. His research interests include computational photography, VLSI for image sensors, and computer vision.
\end{IEEEbiography}

\vspace{-33pt}

\begin{IEEEbiography}[{\includegraphics[width=1in,height=1.25in,clip,keepaspectratio]{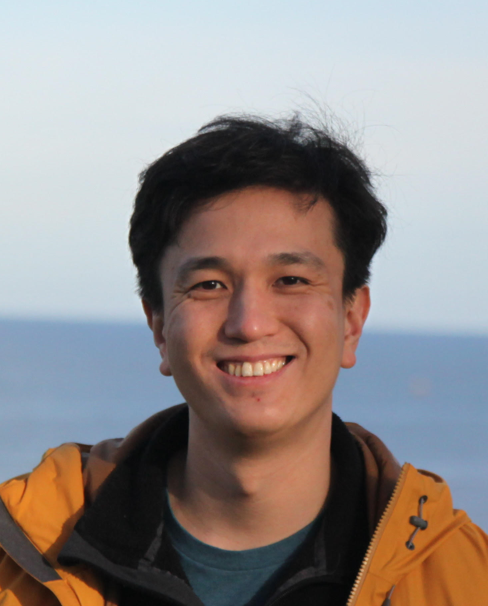}}]{Jie (Jack) Zhang}
received his B.S. and Ph.D degrees in Electrical and Computer Engineering from Johns Hopkins University (JHU) in 2010 and 2016 respectively. He is currently a Research Scientist at the Picower Institute for Learning and Memory and the Department of Brain and Cognitive Sciences at MIT. His research interests include VLSI for biomedical applications, image sensors, electrophysiology, functional neural imaging, and the study of neural mechanisms of spatial navigation and memory consolidation. He also builds open-source tools for the neuroscience community and is a key contributor of Open-Ephys.
\end{IEEEbiography}

\vspace{-33pt}

\begin{IEEEbiography}[{\includegraphics[width=1in,height=1.25in,clip,keepaspectratio]{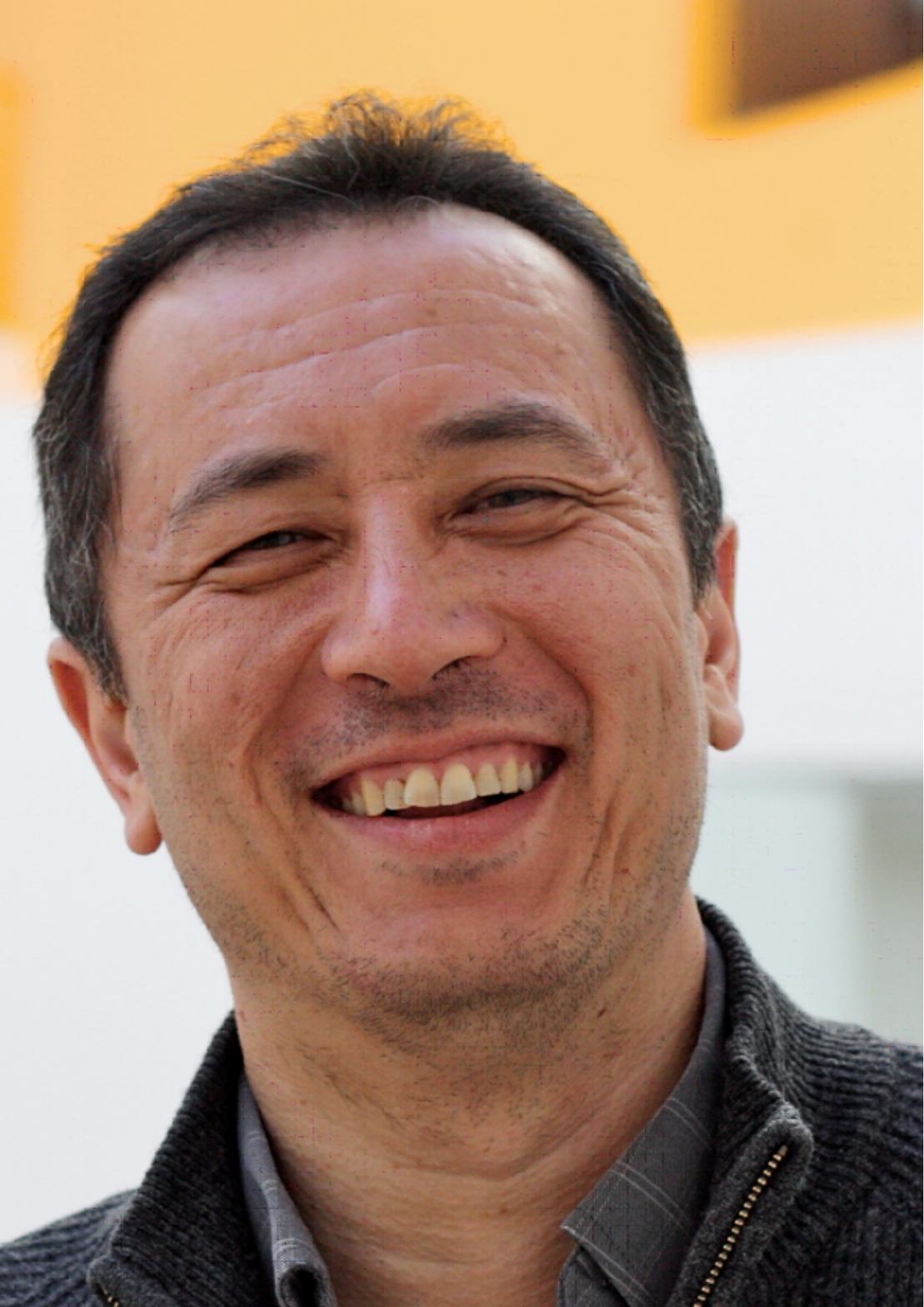}}]{Matthew A. Wilson}
received his BS in Electrical Engineering from Rensselaer Polytechnic Institute, his MS in Electrical Engineering from the University of Wisconsin, Madison, and his PhD. in Computation and Neural Systems from the California Institute of Technology.  He joined the faculty at MIT in the Department of Brain and Cognitive Sciences in 1994 and is the Sherman Fairchild Professor of Neurobiology. He studies the role of sleep in learning and memory, and brain systems that contribute to spatial navigation, and decision-making, and their possible involvement in neurological diseases and disorders through the use of microelectrode arrays implanted in the brains of freely behaving rodents.  He is currently Associate Director of The Picower Institute for Learning and Memory at MIT, and Associate Director of the Center for Brains Minds and Machines
\end{IEEEbiography}

\vspace{-33pt}

\begin{IEEEbiography}[{\includegraphics[width=1in,height=1.25in,clip,keepaspectratio]{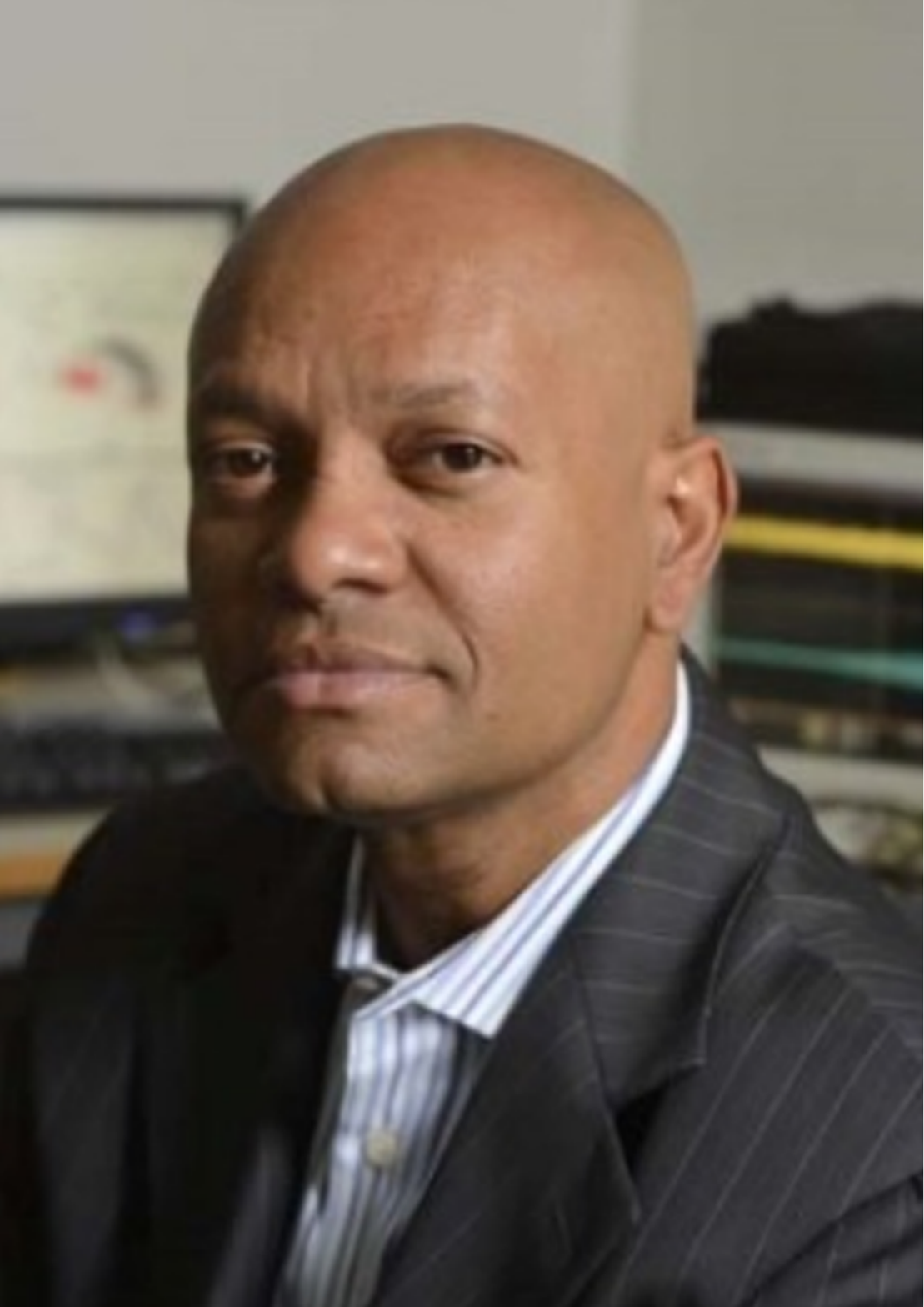}}]{Ralph Etienne-Cummings}
 (F'13) received the B.S. degree in physics from Lincoln University, Lincoln, PA, USA, in 1988, and the M.S.E.E. and Ph.D. degrees in electrical engineering from the University of Pennsylvania, Philadelphia, PA, USA, in 1991 and 1994, respectively. He is currently a Professor of electrical and computer engineering, and computer science with The Johns Hopkins University, Baltimore, MD, USA. He was the Founding Director of the Institute of Neuromorphic Engineering. He has authored more than 200 peer-reviewed articles and holds numerous patents. Dr. Etienne-Cummings has served as Chairman of IEEE Circuits and Systems (CAS) Technical Committees and was elected as a member of CAS Board of Governors. He also serves on numerous editorial boards. He is a recipient of the NSFs Career and Office of Naval Research Young Investigator Program Awards. He was a Visiting African Fellow at the University of Cape Town, Fulbright Fellowship Grantee, Eminent Visiting Scholar at the University of Western Sydney and has also won numerous publication awards, including the 2012 Most Outstanding Paper of the IEEE Transactions on Neural Systems and Rehabilitation Engineering. In addition, he was recently recognized as a Science Maker, an African American history archive.
\end{IEEEbiography}

\vfill

\end{document}